\documentclass[preprint,12pt,longnamesfirst,epsf]{aastex}

\newcommand \kms          {km~s$^{-1}$}

\begin{document}

\title{Hot Interstellar Gas and Stellar Energy Feedback in the 
Antennae Galaxies}
\shorttitle{Hot Gas in Antennae Galaxies}

\author{Joannah M.\ Metz\altaffilmark{1}, Randall L.\ 
Cooper\altaffilmark{1,2},  
Mart\'{\i}n A.\ Guerrero\altaffilmark{1,3}, 
You-Hua Chu\altaffilmark{1}, C.-H.\ Rosie Chen\altaffilmark{1}, 
Robert A.\ Gruendl\altaffilmark{1,4}}

\altaffiltext{1}{Astronomy Department, University of Illinois, 
        1002 W. Green Street, Urbana, IL 61801;
        jmmetz@uiuc.edu, rcooper1@astro.uiuc.edu, mar@astro.uiuc.edu, 
        chu@astro.uiuc.edu, c-chen@astro.uiuc.edu, gruendl@astro.uiuc.edu}
\altaffiltext{2}{Now at Department of Astronomy, Harvard University;
rcooper@cfa.harvard.edu}
\altaffiltext{3}{Now at Instituto de Astrof\'{\i}sica de Andaluc\'{\i}a 
 (CSIC), Spain; mar@iaa.es}
\altaffiltext{4}{Visiting Astronomer, Cerro Tololo Inter-American 
Observatory, National Optical Astronomy Observatories, which are 
operated by the Association of Universities for Research in 
Astronomy, Inc.\ under cooperative agreement with the National 
Science Foundation.}

%\received{}
%\accepted{}
%\journalid{}{}
%\articleid{}{}

\begin{abstract}

We have analyzed $Chandra$ archival observations of the Antennae 
galaxies to study the distribution and physical properties of its 
hot interstellar gas. 
Eleven distinct diffuse X-ray emission regions are selected 
according to their underlying interstellar structures and star 
formation activity.  
The X-ray spectra of these regions are used to determine their 
thermal energy contents and cooling timescales.  
Young star clusters in these regions are also identified and their 
photometric measurements are compared to evolutionary stellar 
population synthesis models to assess their masses and ages.  
The cluster properties are then used to determine the stellar 
wind and supernova energies injected into the ISM.
Comparisons between the thermal energy in the hot ISM and the 
expected stellar energy input show that young star clusters 
are sufficient to power the X-ray-emitting gas in some, but not
all, active star formation regions.
Super-star clusters, with masses $\ge 1\times10^5$ M$_\odot$, heat
the ISM, but the yield of hot interstellar gas  is not directly
proportional to the cluster mass.
Finally, there exist diffuse X-ray emission regions which do not show 
active star formation or massive young star clusters.  
These regions may be powered by field stars or low-mass clusters 
formed within the last $\sim$100 Myr.

\end{abstract}  


\newpage
\section{Introduction}

The interacting galaxies NGC\,4038/4039, nicknamed ``The Antennae''
galaxies, are the nearest example of a major merger of two massive disk 
galaxies and provide an excellent opportunity to study galaxy collisions 
in detail.  
The Antennae are located at a distance of only 19.2 Mpc (for H$_0 $= 75
km s$^{-1}$ Mpc$^{-1}$), allowing a detailed examination of astrophysical
processes with high resolution at multiple wavelengths 
\citep{Whitmore99,ZFW01}.

The gravitational interactions from the merger trigger star formation and
produce conditions appropriate for globular clusters to form.  Thus, the
Antennae contain many sites of active star formation, giant \ion{H}{2}
regions, and young super-star clusters.  Massive stars interact with the
interstellar medium (ISM) via fast stellar winds and supernova
ejecta.  These dynamic interactions produce hot ($\ge10^6$ K) ionized gas
which can be studied in X-rays.  Examining the characteristics of the
X-ray emission from the Antennae galaxies is important in determining how
the hot interstellar gas is energized and distributed.  This will allow 
us to study stellar energy feedback processes in different types of
star formation regions, especially in young globular clusters, i.e.,
young super-star clusters, which are not available in our Galaxy or 
nearby Local Group galaxies.

Many X-ray telescopes have observed the Antennae galaxies.  
The \emph{Einstein} observations were the first to show diffuse X-ray
emission, but were unable to resolve the point sources from the diffuse 
emission \citep{Fabbiano82, Fabbiano83}.  
Subsequent observations with \emph{ROSAT} PSPC, which had higher 
spatial and spectral resolution, showed 55\% of the X-ray flux 
(in the PSPC band of 0.1-2.4 keV) from the Antennae to be resolved 
into discrete sources with the rest predominantly diffuse emission 
\citep{Read95}.  
\emph{ROSAT} HRI observations were able to resolve twelve discrete 
sources \citep{Fabbiano97}.
\emph{ASCA} observations confirmed a hard X-ray component and softer
gaseous emission \citep{Read95,Sansom96}. 
The advent of the \emph{Chandra X-ray Observatory} finally made it 
possible to clearly resolve point sources from diffuse emission.
Using \emph{Chandra} ACIS-S observations, 48 point X-ray sources are 
detected in the Antennae galaxies; these point sources contribute  
81\% of the emission in the 2.0--10.0 keV band but only 32\% of the 
emission in the 0.1--2.0 keV band \citep{Fabbiano01}.
The discrete sources in the Antennae have diverse spectral properties, 
with the most luminous ones showing hard emission consistent with X-ray
binaries and the faintest ones displaying softer emission similar to 
supernova remnants (SNRs) or hot ISM \citep{Zezas02a}.
Many X-ray sources show good correspondence with the young stellar 
clusters \citep{Zezas02b}.

\emph{Chandra} observations of the Antennae galaxies vividly portray
a dynamic ISM with wide-spread hot gas throughout both galaxies
\citep{Fabbiano01}.
The active star cluster formation that is responsible for the large
\ion{H}{2} complexes, such as giant \ion{H}{2} regions and supergiant
shells, must also contribute to the production of hot ISM.
Previous investigations of hot gas in the Antennae did not take into
account the physical association with the young stellar clusters
\citep[e.g.,][]{Fetal03}.
Therefore, we have compared the distribution of diffuse X-ray 
emission with the underlying star cluster content and cooler 
interstellar structures revealed in optical images, identified 
physical associations between them, and analyzed the physical 
properties of the hot interstellar gas
accordingly.  We have further used the photometry of the underlying
clusters \citep{Whitmore99} to determine their ages and masses 
in order to derive the expected stellar wind and supernova energies 
injected into the ISM.
These results allow us to compare the thermal energy stored in the 
hot X-ray-emitting medium to the stellar energy input and compare the 
cooling timescale of the hot gas to the age of the clusters so that 
we may assess the role of the young clusters in energizing the ISM.

In this paper, we describe the \emph{Chandra} observations and
complementary \emph{Hubble Space Telescope} images and ground-based 
echelle observations in \S 2, report the identification of distinct 
interstellar diffuse X-ray sources in \S 3, and analyze physical 
properties of the hot, X-ray-emitting ISM in \S 4.
We consider the stellar energy feedback from the clusters and compare
it to the thermal energy stored in the hot ISM in \S 5, and discuss the
production of hot ISM with and without super-star clusters in \S 6.
A summary is given in \S 7.

\section{Observations and Reduction}

The Antennae galaxies were observed with the back-illuminated CCD 
chip S3 of the ACIS-S array onboard the \emph{Chandra X-Ray Observatory} 
on 1999 December 1 for a total exposure time of 72.5 ks (Observation
ID: 315; PI: Murray), on 2001 December 29 for 69.9 ks (Observation
ID: 3040; PI: Fabbiano), and on 2002 April 18 for 68.0 ks (Observation
ID: 3043; PI: Fabbiano). 
We retrieved level 1 and level 2 processed data from the \emph{Chandra}
Data Center.  The data reduction and analysis were performed using the
\emph{Chandra} X-Ray Center software CIAO v2.2.1 and HEASARC FTOOLS and
XSPEC v11.0.1 routines \citep{Arnaud96}.
The latter two observations were affected by high background for 11.8 ks
and 14.6 ks, respectively; these high-background periods were excluded 
in our analysis. 
The three observations were merged to increase the signal-to-noise 
ratio for the analysis of the spatial distribution and spectral 
properties of the X-ray emission.  
The spectral analysis was carried out by computing the calibration 
files for the individual observations and averaging them using weights 
proportional to their net exposure times.

To associate the X-ray emission of the Antennae galaxies to the cooler
ionized ISM and the underlying stellar content, we have also retrieved
archival \emph{Hubble Space Telescope} images taken with the Wide Field 
and Planetary Camera 2 (WFPC2) in the \emph{F555W} and \emph{F658N} filters.
The \emph{F555W} image, similar to a $V$-band image, has a total 
exposure of 4,400 s and shows the star clusters.  
The {\emph{F658N} filter, originally designed as an [\ion{N}{2}] filter, 
contains the red-shifted H$\alpha$ line over the velocity range covered
by the Antennae galaxies.
Thus, the \emph{F658N} image of the Antennae is essentially an H$\alpha$
image contaminated only by the weak [\ion{N}{2}] $\lambda$6548 line,
and should show the 10$^4$ K ionized ISM well.

To study the dynamics of a supergiant shell and three giant \ion{H}{2}
regions with bright X-ray emission, we obtained high-dispersion 
spectroscopic observations with the echelle spectrograph on the Blanco 
4m telescope at the Cerro Tololo Inter-American Observatory on 2002 
March 24 and 25.  
For these observations, a 79 line mm$^{-1}$ echelle grating was used.
The spectrograph was configured in a long-slit, single-order mode
by replacing the cross-disperser by a flat mirror and inserting a 
post-slit broad H$\alpha$ filter (central wavelength 6563 \AA\ with 
75 \AA\ FWHM).
The long-focus red camera was used to obtain a reciprocal dispersion of 
3.5 \AA\ mm$^{-1}$ at H$\alpha$.  The SITe2K \#5 CCD detector with 
24 $\mu$m pixel size was read out with a 2-pixel binning in the spatial
direction, providing a spatial sampling of 0\farcs 53 pixel$^{-1}$ 
along the slit and a spectral sampling of $\sim$0.08 \AA\ pixel$^{-1}$ 
along the dispersion axis.

A 1\farcs5 wide slit was used in the echelle observations.  
A total of four long-slit observations were obtained with 1,200 sec 
integration time each.  
Two of the observations were centered at a single position,
$\alpha$(J2000)$=$12$^{\rm h}$01$^{\rm m}$51\rlap{.}{$^{\rm s}$}3,
$\delta$(J2000)$=$$-$18$^\circ$51\arcmin47\arcsec, with the slit 
oriented in the east-west direction.  The other two 
observations were made with the slit oriented in the north-south
direction centered at the positions:
$\alpha$(J2000)$=$12$^{\rm h}$01$^{\rm m}$50\rlap{.}{$^{\rm s}$}3, 
$\delta$(J2000)$=$$-$18$^\circ$52\arcmin14\arcsec, and 
$\alpha$(J2000)$=$12$^{\rm h}$01$^{\rm m}$50\rlap{.}{$^{\rm s}$}6, 
$\delta$(J2000)$=$$-$18$^\circ$52\arcmin14\arcsec.
These three slit positions will be called S1, S2, and S3, respectively.
The echelle observations were reduced using standard packages in
IRAF\footnote{IRAF is distributed by the National Optical Astronomy
Observatories, which are operated by the Association of Universities 
for Research in Astronomy, Inc., under cooperative agreement with the 
National Science Foundation.}.  
All spectra were bias and dark corrected, and cosmic-ray events were 
removed by hand.  
Observations of a Th--Ar lamp were used to determine the dispersion 
correction and then telluric lines in the source observations were 
used to check the absolute wavelength calibration.
The resultant spectra had an instrumental resolution of 0.26 \AA\ (12 
\kms\ at H$\alpha$), as measured by the FWHM of the unresolved telluric 
emission lines.

\section{Interstellar Sources of X-ray Emission from the Antennae}

The X-ray emission from the Antennae galaxies has a complex 
morphology, consisting of both point sources and diffuse emission.  
To search for X-ray emission from the hot interstellar gas, we have 
made adaptively smoothed images of the Antennae in the soft 
(0.3--2.0 keV) and hard (2.5--6.0 keV) bands and present these 
images in Figures 1a and 1b.
Comparisons between these two images show that the diffuse 
emission is soft and some small, discrete sources are also soft.
The diffuse soft X-ray emission most likely originates from the
hot ($>$ $10^6$ K) ISM.  The discrete soft X-ray sources are 
probably also from the hot ISM and may be unresolved young SNRs.
The heating sources of the hot ISM and the progenitors of the 
supernovae are both likely massive stars.

To examine the relationship between the hot ISM and the massive 
star formation regions, we have plotted contours showing the soft 
X-ray emission over the H$\alpha$ image of the Antennae in Figure~1c.
A tight correlation is evident between some diffuse X-ray emission 
and the star forming regions, indicating that the massive stars are 
indeed responsible for heating the hot ISM.
However, there are also extended X-ray emission regions which do
not exhibit active star formation.
To determine whether this diffuse X-ray emission may be powered by
previous star formation activity, we present an \emph{HST} WFPC2 
\emph{F555W} image in Figure~1d with young (age $<$ 30 Myr) clusters 
marked by white crosses in filled circles and old (0.25--1.0 Gyr) 
clusters by open circles.
It is clear that bright diffuse X-ray emission is more strongly 
correlated with the young clusters than the old clusters.

In regions where diffuse X-ray emission is correlated with recent star 
formation activity, two types of interstellar structures are easily
resolved: giant \ion{H}{2} regions and supergiant shells.
Giant \ion{H}{2} regions with sizes of $\ge$ 100 pc are associated 
with young star formation regions where massive stars 
with powerful ionizing fluxes are prevalent.
Some giant \ion{H}{2} regions are located close to the galactic nuclei 
and may be associated with nuclear starbursts; these will be referred
to as circumnuclear starburst regions. 
Supergiant shells with sizes approaching 1,000 pc have dynamic ages 
greater than a few $\times 10^7$ yr and are associated with older star 
formation activity.
By comparing the physical properties of the hot ISM associated with
giant \ion{H}{2} regions and supergiant shells, it is possible to
study the content and evolution of the hot ISM associated with star
formation.

We have therefore chosen for detailed analyses diffuse X-ray sources 
associated with (1) circumnuclear starburst regions, (2) giant 
\ion{H}{2} regions, (3) supergiant shells, and (4) no obvious recent
star formation activity.
The 11 source regions we have selected are marked in Figure~1a
and numbered counterclockwise starting from the west side 
of the galaxies.
Elliptical source apertures are used to extract X-ray fluxes and 
spectra from these regions.
Whenever source regions contain point sources, small regions around 
the point sources are excised in the analysis of the diffuse X-ray 
emission.
Table~1 lists the positions, major and minor axes, position angles 
of the major axes, volumes of these source regions, and the total
background-subtracted counts detected in the three observations.   
The volume is calculated assuming an ellipsoidal geometry with the
third axis equal to the minor axis of the elliptical region.
The excised regions around point sources are given in the notes
for Table~1.
Figure~2 shows a detailed view of the selected X-ray-emitting
regions; in each panel, an H$\alpha$ image is presented with 
overlays of soft X-ray (0.3--2.0 keV) contours, outlines of the 
elliptical source apertures, and the identifications of underlying 
young clusters.
Below we describe the 11 source regions, grouped according to their 
associated star formation activity or interstellar structures.

\subsection{Circumnuclear Starburst Regions (D04 and D05)}

The nuclei of the two galaxies NGC 4038 and NGC 4039 can be identified 
from their high CO concentrations \citep[e.g.,][]{Letal01} or the red
color in a true-color image \citep{Whitmore99}.
Both nuclei are associated with active star formation and bright X-ray 
emission.
Due to the presence of bright hard X-ray emission in the nucleus of 
NGC 4039, we excluded this region from our study.
The nucleus of NGC 4038 is bisected by a dust lane, thus two giant 
\ion{H}{2} regions are observed.
The H$\alpha$ morphologies of these two giant \ion{H}{2} regions are 
quite different, suggesting that they may be independent from each 
other and should be examined separately.  
Therefore, we defined them as two separate regions.

Regions D04 and D05 are the two source regions associated with the 
nucleus of NGC 4038.
Each contains a giant \ion{H}{2} region and two young super-star clusters. 
Each region also contains a distinct point source, which is excised in 
the analysis of the diffuse X-ray emission.
The giant \ion{H}{2} region in D05 has a shell structure measuring 
250 pc across. 

\subsection{Giant \ion{H}{2} Regions (D01, D02, D07, D09, and D10)}

We chose these source regions based on the coincidence between 
diffuse X-ray emission and giant \ion{H}{2} regions.
The size of a source region is selected to include all the
enhanced diffuse X-ray emission.
Inspection of the H$\alpha$ image shows that each of the source regions 
contains a variety of interstellar structures.
Some contain multiple giant \ion{H}{2} regions with distinct 
morphologies, and some contain large shell 
structures extending from the giant \ion{H}{2} regions.
The five that contain giant \ion{H}{2} regions are described below. 
 
Region D01
contains three prominent giant \ion{H}{2} regions with sizes of 280 pc, 
230 pc, and 280 pc as well as a large shell structure measuring almost 
500 pc across.
The diffuse X-ray emission peaks in the space between the three giant 
\ion{H}{2} regions, including the large shell structure.  
It also contains six young clusters  ($\le$ 30 Myr) coinciding with 
the star formation regions.  

Region D02 
contains a giant \ion{H}{2} region measuring 190 pc across and a large 
shell structure, 400 pc in size, extending to the northwest.
The diffuse X-ray emission peaks at the giant \ion{H}{2} region.
Three young clusters are associated with the brightest star formation
region. 

Region D07 
contains a string of three giant \ion{H}{2} regions measuring 280 pc, 
230 pc, and 120 pc across.  
An X-ray point source is contained in region D07, but it was excised 
when we analyzed the diffuse emission.  
It also contains four young clusters distributed along the star 
formation regions.

Region D09 contains one giant \ion{H}{2} region and two young clusters.
The associated X-ray emission peaks at the giant \ion{H}{2} region.
A point source is detected to the northwest of the giant \ion{H}{2} 
region; this point source has been excised from the diffuse X-ray
emission.

Region D10 
contains a star forming region and a large shell structure measuring 
650 pc across, approaching the size of a supergiant shell.  
The associated X-ray emission peaks at the giant \ion{H}{2} region and
extends to the southwest.
D10 contains eight young clusters associated with the giant \ion{H}{2}
region; four of these clusters are massive, super-star clusters.
Two point sources are detected within this source region; both point
sources are excised from the diffuse X-ray emission.

\subsection{Supergiant shell (D03)}

Region D03 was selected to coincide with a supergiant shell measuring 
more than 900 pc across.  
Supergiant shells with this large size are expected to break out of
the gaseous disk of a galaxy, but the X-ray emission outside the D03
supergiant shell quickly drops off, indicating either no breakout or 
a breakout along the line of sight. 
This region contains diffuse X-ray emission centered on the
supergiant shell.  It also contains three young clusters.

\subsection{Regions without Active Star Formation (D06, D08, and D11)}

Three bright diffuse X-ray emitting regions that do not contain obvious
star formation activity to power the emission were also chosen for study.
The origin of the hot gas in these regions may be different than the 
origins of those discussed earlier.

Region D06 
is partially surrounded by some \ion{H}{2} regions, including 
those in regions D04 and D07, but has no active star formation
within the diffuse X-ray emitting region.
Furthermore, it is not located on a spiral arm.     
D06 contains one young cluster displaced from the peak of the X-ray 
emission.

D08 is another diffuse emission region without corresponding star 
formation activity.  
It is adjacent to some regions of star formation activity along its 
southwest quadrant.
It contains two young clusters in the northwest part of the region.  

The region D11 is sandwiched between the nucleus of NGC 4039 and a 
prominent star forming region associated with our source region D10.  
No clusters or star formation activity are observed in region D11.

\section{Physical Properties of the Hot ISM}

The diffuse X-ray emission originates from the hot ISM.
By modeling the X-ray fluxes and spectra, we can determine
the plasma temperature, volume emission measure, and 
foreground absorption of this hot ISM.
These results allow us to further estimate the thermal energy 
and cooling time of the hot gas.

\subsection{Spectral Analysis and Results}

To extract X-ray spectra from the 11 diffuse X-ray sources defined 
in \S3, we have used the elliptical source regions given in Table~1 
and shown in Figure 1a, and an annular background region centered 
on the Antennae galaxies with inner and outer radii of 130\arcsec\ 
and 175\arcsec, respectively.  
The background spectrum is then scaled by the surface area for each 
source region and subtracted.  
Because of the high surface brightnesses of our diffuse sources,
the background correction is small in every case, $\leq4\%$ the 
number of counts in the raw spectrum and much less than the 
1-$\sigma$ uncertainty plotted for each spectral channel.    
As noted in Table~1, circular regions of radii 1\farcs2 -- 1\farcs7 
centered on the point sources are used to exclude their X-ray 
emission from our analysis of the diffuse sources.
The background-subtracted, point-source-excised spectra extracted 
from these regions are then modeled and analyzed.

We have modeled the X-ray spectra using a single thermal plasma
component with a fixed solar abundance and a fixed foreground
absorption column density, approximated by the Galactic \ion{H}{1}
column density $3.9\times10^{20}$ cm$^{-2}$.
The rationale for our choice of this model is given in Appendix A,
and comparisons among different models and their effects on
the thermal energy are given in Appendix B.
Only the spectral range of 0.5--2.0 keV is used for the spectral 
fitting, since there are no measurable counts above 2.0 keV.
The best-fit models are plotted over the observed spectra in Figure 3.
The results of the spectral fitting are presented in Table~2: 
column (1) is the region name; columns (2) and (3) are the 
plasma temperatures $T$ in 10$^{6}$ K and $kT$ in keV, with 3$\sigma$ 
errors; column (4) is the normalization factor 
$A = \frac{3.1\times10^{4}}{4{\pi}d^2}{\int}N_{\rm e}^2dV$ cm$^{-5}$, 
where $d$ is the distance in pc, $N_{\rm e}$ is the electron density 
in cm$^{-3}$, and $V$ is the volume of the X-ray-emitting gas in pc$^3$; 
column (5) is the rms $N_{\rm e}$ derived later in $\S$4.2; 
column (6) is the observed X-ray flux in the 0.5--2.0 keV energy band; 
and column (7) is the X-ray luminosity in the same energy band for a 
distance of 19.2 Mpc.
These diffuse X-ray emission regions typically have temperatures of 
4 -- $9 \times 10^6$ K, and X-ray luminosities of 
$3 \times 10^{38}$ -- $2 \times 10^{39}$ ergs s$^{-1}$ in the 
0.5--2.0 keV energy band.

\subsection{Thermal Energy Content and Cooling Time of the Hot Gas}

The thermal energy of the hot gas in an X-ray-emitting region is
$ E_{\rm th} = \frac{3}{2} k T N \epsilon V$,
where $T$ is the plasma temperature from the best spectral fit as
given in Table 2, $N$ is the total particle number density, $\epsilon$ 
is the volume filling factor, and $V$ is the volume given in Table 1.
The total particle number density $N$, approximated by the sum of 
electron density, hydrogen density, and helium density, is 
$N_{\rm e}+N_{\rm H}+N_{\rm He}~\sim~1.92 N_{\rm e}$, assuming
a canonical $N_{\rm He}/N_{\rm H}$ ratio of 0.1.
For a given volume filling factor $\epsilon$, the rms $N_e$ can be 
derived from the normalization factor $A$ of the best spectral fit 
given in Table~2: $N_{\rm e}$ = $0.02~d ~(A/\epsilon V)^{1/2}$ 
cm$^{-3}$, with $d$ in pc, $A$ in cm$^{-5}$, and $V$ in pc$^3$.
The rms $N_{\rm e}$ and thermal energy are dependent on the filling 
factor, $N_{\rm e} \propto \epsilon^{-1/2}$ and 
$E_{\rm th} \propto \epsilon^{1/2}$.
The filling factor $\epsilon$ is most likely between 0.5 and 1.0.
We have assumed a filling factor of 1.0, and we present the rms $N_e$
in Table 2 and the thermal energy $E_{\rm th}$ in Table 3.
The rms $N_{\rm e}$ is typically 0.02--0.08 cm$^{-3}$, and the 
thermal energy ranges from $4\times10^{53}$ ergs~s$^{-1}$ to
$5\times10^{54}$ ergs~s$^{-1}$.

The cooling timescale of the hot interstellar gas $t_{\rm cool}$ is
$\sim {(3/2) N k T}/\Lambda(T)$,
where $\Lambda(T)$ is the cooling function for interstellar 
gas at temperature $T$.  
Since the plasma temperatures of all our diffuse X-ray
sources are a few $\times 10^{6}$ K, we adopt a constant value of 
$\Lambda(T)/N_{\rm H}^2 = 2.5 \times 10^{-23}$ ergs cm$^{3}$ s$^{-1}$ 
\citep{DM72}.  
As $N_{\rm e} \sim 1.2 N_{\rm H}$ and $N \sim 1.92 N_{\rm e}$, the 
cooling timescale $t_{\rm cool}$ becomes $\sim {0.73\, T}/N_{\rm e}$ yr,
with $T$ in K and $N_{\rm e}$ in cm$^{-3}$.
The cooling timescales calculated for the 11 source regions,
typically several tens of Myr, are listed in Table 3.

\section{Stellar Energy Feedback and Energy Balance}

As noted previously and listed in Table 3, all of our diffuse X-ray
emission regions except D11 contain at least one young star cluster.
Multi-band photometric measurements of a cluster can be used to 
estimate the mass and age of the cluster, which in turn allow us
to assess the stellar energy injected into the ISM via fast stellar 
winds and supernova explosions. 
In this section we compare the thermal energy in the hot ISM
with the stellar energy feedback and examine the energy balance.

\subsection{Cluster Masses and Ages}

To simulate the photometric measurements of clusters in the 
Antennae galaxies, we use the Starburst99 code by \citet{Letal99}.
We have chosen solar metallicity for the clusters and a Salpeter 
initial mass function \citep{S55} with respective lower and 
upper mass limits of 1 M$_{\odot}$ and 100 M$_{\odot}$. 
$UBVI$ magnitudes of clusters with initial masses of $5\times10^4$,
$1\times10^5$, and $5\times10^5$ M$_\odot$ are generated using
Starburst99 for ages from 1 Myr to 30 Myr.
As the extinctions of the clusters are unknown, we follow
\citet{Whitmore99} and adopt the reddening-free colors
$Q_1 = (U - B) - 0.72(B - V)$ and $Q_3 = (U - B) - 0.58(V - I)$.
We present the extinction-free color-color diagram of $Q_1$ vs.\
$Q_3$ in Figure 4a and compare the locations of the 50 brightest 
young clusters \citep[photometry from Table 1 of][]{Whitmore99} to 
the theoretical evolutionary tracks to determine the cluster ages.
We further adopt the extinction-free magnitude $W = M_V - R_V (B - V)$ 
\citep{M82}, where $R_V = A_V/E(B-V) = 3.2$, and produce extinction-free 
color-magnitude diagrams of $W$ vs.\ $Q_1$ and $W$ vs.\ $Q_3$.
In Figure 4b, we illustrate the evolutionary tracks in such
color-magnitude diagrams for a cluster with an initial mass of
$1\times10^5$ M$_\odot$ from 1 Myr to 30 Myr.
In Figures 4c \& d, we compare the observed locations of the clusters
in the extinction-free color-magnitude diagrams to the evolutionary 
tracks of clusters with different initial masses and assess the
initial masses of the clusters based on the above age estimates.

The initial masses and ages of the 50 brightest young clusters in the
Antennae galaxies are summarized in Table 4.  
We caution that systematic errors are likely to exist, as there
appear to be no clusters in the age range of 8--13 Myr, and many
clusters occupy positions that are at $\Delta Q$ = 0.1 -- 0.2 mag
from the theoretical evolutionary tracks in the $Q_1$ vs.\ $Q_3$ 
diagram.
Uncertainties in the Starburst99 modeling and errors in the 
photometric measurements both contribute to the large discrepancy
between observations and expectations.
The masses and ages in Table 4 should therefore be viewed as crude, 
order-of-magnitude approximations.

We have also explored the evolutionary stellar population synthesis 
models by \citet{BC03}.  The synthetic photometries of clusters at 
different ages produced by Bruzual \& Charlot's models differ 
significantly from that produced by Starburst99.
These differences are attributed to the uncertainties in theoretical 
models of massive star evolution by the Geneva and Padova groups
used by Starburst99 and Bruzual \& Charlot's models, respectively.
We have chosen Starburst99 because it also computes the stellar energy
output in the form of fast wind and supernova explosion, which will be
discussed next.

\subsection{Stellar Wind Energy}

Stellar winds from the star clusters in the diffuse X-ray emission
regions contribute to the overall stellar energy feedback.  
To establish an upper bound to the stellar wind's contribution, 
we estimate the average maximum stellar wind energy of main 
sequence O and B stars.  To do so, we make two simplifying 
assumptions.  First, we assume that half of the mass of a star 
is ejected as stellar wind.  Second, we assume that all of 
the stellar wind is ejected at the 
terminal velocity as given by Prinja, Barlow, \& Howarth (1990).  
They define the stellar wind terminal velocity as the velocity of 
outflowing matter that is sufficiently far from the star such that 
it experiences a negligible gravitational force, but has not yet 
significantly interacted with the ISM. 

Both of these assumptions overestimate the stellar wind energy.  
Typically the fraction of a star's mass ejected as stellar wind is 
less than one half of its initial mass.  
When a massive star leaves the main sequence, its radius 
increases substantially.  
Consequently, its escape velocity decreases.  
Stellar wind velocity is usually a few times the escape velocity, so 
the stellar wind velocity at this evolutionary stage is notably 
lower than that used in our calculation.

The upper bound of the stellar wind energy is 
$E_{\rm wind} = \frac{1}{4}Mv_{\infty}^{2}$, where $M$ 
is the initial mass of the star and $v_{\infty}$ is the terminal 
velocity of the stellar wind.
The weighted average of $E_{\rm wind}$, assuming a Salpeter initial 
mass function, is thus
\begin{equation}
\overline{E}_{\rm wind} = \frac{\sum_{i} E_{i}M_{i}^{-2.35}}{\sum_{i}
 M_{i}^{-2.35}}.
\end{equation}
The masses and terminal wind velocities of stars with a range of 
spectral types are given in Table 5; using these, we calculate 
$\overline{E}_{\rm wind} = 7 \times 10^{49}$ ergs.  
We assume that only stars with masses greater than 10 M$_{\odot}$ 
contribute significantly to the stellar wind energy.  
Stars with masses less than 10 M$_{\odot}$ have very low mass loss 
rates and stellar wind velocities.  
Thus the number of stars in a cluster that contribute to the stellar 
wind energy input is 
\begin{equation}
N_{\rm wind} = M_{\rm cluster} \frac{\int_{10 {\rm M}_{\odot}}^{100
 {\rm M}_{\odot}} 
M^{-2.35}\,\mathrm{d}M}{\int_{1 {\rm M}_{\odot}}^{100 {\rm 
M}_{\odot}}M^{-1.35}\,
\mathrm{d}M} \hspace{0.25cm},
\end{equation}
where $M_{\rm cluster}$ is the mass of the cluster determined in \S5.1.
The upper bound for the total stellar wind energy, 
$E_{\rm wind} = \overline{E}N_{\rm wind}$, is listed in column 7 of 
Table 4.

We have also calculated the stellar wind energy using Starburst99.  The 
results are given in column 9 of Table 4.
Most of these values 
are $\sim$ 3 times our estimated upper bounds of the stellar 
wind energy.  The difference most likely arises from the high terminal 
wind velocities used in Starburst99 \citep{LRD92}.

\subsection{Supernova Energy}

For a cluster with a given mass and age, the number of stars that have 
exploded as supernovae can be calculated.
A star of mass $M < 30$ M$_{\odot}$ will remain on the main sequence
for approximately $\tau \sim  10(\frac{M}{{\rm M}_{\odot}})^{-2.5}$ Gyr 
\citep{BM98},
so all stars with $M > M_{min} = (\frac {10\,\mathrm{Gyr}}{\tau})^{0.4}$
M$_\odot$ will have already exploded. 
The number of stars that have exploded as supernovae is
\begin{equation}
N_{\rm SN} = M_{\rm cluster} \frac{\int_{M_{\rm min}}^{100 {\rm M}_{\odot}} 
M^{-2.35}\,\mathrm{d}M}{\int_{1 {\rm M}_{\odot}}^{100 {\rm 
M}_{\odot}}M^{-1.35}\,\mathrm{d}M} \hspace{0.25cm},
\end{equation}
Each supernova injects roughly $10^{51}$ ergs of explosion energy into 
the ISM.
Therefore the total supernova energy input from the  
cluster is approximately $E_{\rm SN} = N_{\rm SN} \times 10^{51}$ ergs.
$M_{\rm min}$,  $N_{\rm SN}$, and $E_{\rm SN}$ are given in columns 
4--6 of Table 4.

We have also calculated the supernova energy input using Starburst99, 
and these values are listed in column 8 of Table 4.
The Starburst99 estimates are consistent with our estimates for clusters 
older than $\sim$ 10 Myrs.
However, our approximation overestimates the number of supernovae for 
younger clusters, especially those with ages less than 3.5 Myrs.
 For clusters between 3.5 Myrs and 10 Myrs old, our estimated 
$E_{\rm SN}$'s are usually within a factor of two higher than the values 
estimated by Starburst99.

\subsection{Comparison of Stellar Energy Input to Thermal Energy}

Table 3 compares the stellar energy input from young star clusters
and the thermal energy in each of the 11 diffuse X-ray emission
regions: column 1 gives the region names; column 2 lists the young
star clusters within each region, with super-star clusters ($\ge
1\times10^5$ M$_\odot$) highlighted in boldface; column 3 gives
the thermal energy in the hot gas; and columns 4 and 5 list the
stellar energy ($E_\star$) inputs, i.e., the sum of stellar wind and 
supernova energies, from the young clusters encompassed in each region
estimated using our method outlined above and using Starburst99, 
respectively.
It can be seen that in regions with recent star formation activity,
the stellar energy input from the young clusters is either comparable
to or a few times greater than the observed thermal energy in the 
X-ray-emitting gas.
Note that stellar energy is usually converted to both thermal and 
kinetic energies of the ambient ISM, and the kinetic energy is 
comparable to or larger than the thermal energy \citep{Ketal98,Cetal04}.
Therefore, in regions D03, D04, D05 and D10, where the stellar energy 
input is several times the thermal energy in the hot gas, the 
underlying young clusters are capable of producing the hot gas.
In regions D01, D02, D07, and D09, where stellar energy input is
only comparable to or just twice as much as the thermal energy,
the young clusters may be insufficient to produce the hot gas.
Three diffuse X-ray emission regions have no obvious star formation
activity, as indicated by the lack of bright \ion{H}{2} regions.
Among these three, D06 and D08 have stellar energy inputs many times
lower than the thermal energies, and D11 contains no young clusters
at all; clearly the hot gas in these regions has not been heated
by young clusters in the last 20 Myr.
These conclusions hold even if the uncertainties in the spectral
fits, as detailed in Appendix B, are taken into account.

Table 3 also lists the cooling timescales of the hot gas in column 6.
These cooling timescales are 55--240 Myr, much greater than the ages 
of the underlying young clusters, $<$ 20 Myr.
As the interstellar gas heated by the clusters has not cooled 
significantly, it is understandable that the temperatures of the
hot gas do not display large variations among regions associated
with giant \ion{H}{2} regions and supergiant shells.

\section{Production of Hot ISM with and without Super-star Clusters}

The Antennae galaxies contain a large number of clusters, including
super-star clusters with masses greater than 10$^5$ M$_\odot$.
One motivation for this work is to investigate the role of super-star 
clusters in the heating of the ISM.
We have compared the locations of the young super-star clusters
from our Table 4 with the distribution of diffuse soft X-ray 
emission and found that every one of them is projected within 
regions of diffuse X-ray emission.
However, not all bright diffuse X-ray emission is associated with
super-star clusters; of the 11 regions selected for this study, 
only eight contain at least one super-star cluster.
Below we compare the diffuse X-ray regions in the Antennae galaxies
with similar regions in other galaxies which do not contain super-star 
clusters.
We will also discuss the origin of hot gas in the three diffuse
X-ray regions, D06, D08, and D11, which lack super-star clusters
and recent active star formation.

\subsection{Hot Gas in Giant \ion{H}{2} Regions and Supergiant Shells}

Seven diffuse X-ray emission regions in the Antennae selected for 
further analysis are characterized by intense star formation activity, 
such as giant \ion{H}{2} regions or circumnuclear starburst regions. 
These regions can be compared to the giant \ion{H}{2} regions in
the spiral galaxy M101 at 7.2 Mpc \citep{WC95,Ketal03}.
The three most luminous giant \ion{H}{2} regions in M101 (NGC\,5461,
NGC\,5462, and NGC\,5471) have X-ray luminosities of $\sim
2\times10^{38}$ ergs~s$^{-1}$ \citep{WC95}.\footnote{A factor of 
(7.2/6)$^2$ is applied to the X-ray luminosities reported by 
\citet{WC95}, as they adopted a distance of 6 Mpc.}
Most of the above seven diffuse X-ray emission regions in the Antennae
are only factors of 2--4 more luminous than the giant \ion{H}{2} regions
in M101.
The star clusters in the three M101 giant \ion{H}{2} regions are all
comparable to or less massive than the R136 cluster, whose mass is 
$\sim 2\times10^4$ M$_\odot$ \citep{CCJ03},
and the aggregate mass of the star clusters in each M101 giant \ion{H}{2}
region is at most comparable to that of a single super-star cluster.
Comparing the masses and X-ray luminosities between the emission regions 
in the Antennae and the M101 giant \ion{H}{2} regions, we find that the 
clusters in the Antennae are not as efficient as the M101 clusters at 
producing diffuse X-ray luminosity.
Most notably, region D10 contains four super-star clusters and four 
R136-class clusters, yet its X-ray luminosity is only a factor of 4
higher than those of the M101 giant \ion{H}{2} regions. 
We therefore conclude that super-star clusters play a significant role in
heating the ISM, but the yield of hot interstellar gas is not directly 
proportional to the cluster mass.
The spatial distribution and physical conditions of the ambient ISM must
play an important role in determining how the stellar winds and supernova
ejecta interact with the ambient medium and what fraction of the stellar
energy input is retained as the thermal energy of hot interstellar gas.

Active star formation regions often contain large interstellar
shell structures; for example, region D01 contains a large shell 
structure extending from the two southern giant \ion{H}{2} regions 
to the east with a diameter of $\sim$ 500 pc.
This large shell is detected in our echelle observation at slit
position S2 (see Figure 5).
Its expansion velocity, $\sim65$ km~s$^{-1}$, is larger than 
most supershells of comparable sizes in dwarf galaxies \citep{M98}.
Six clusters exist in region D01, including a super-star cluster.
Note that the ionized gas near the super-star cluster does not 
expand as fast as the supershell. 
Evidently, the super-star cluster is either in a denser ISM or
has not injected sufficient energy to accelerate the ambient ISM.

It is interesting to note that the largest expansion velocities 
($>$100 km~s$^{-1}$) we detected in D01 are in regions outside 
the main body of the giant \ion{H}{2} regions.  
See the echellograms at S2 and S3 in Figure~5.  
These high-velocity features are most likely associated with SNRs.  
Indeed, one high-velocity feature along S3, marked in Figure~5,
is coincident with an unresolved soft X-ray source marked in 
Figure~1a as a SNR candidate.  

Prolonged and spatially extended star formation activity may produce
supergiant shells with sizes approaching 1,000 pc.
Region D03 contains a supergiant shell which may be compared to LMC-2,
the brightest supergiant shell in the Large Magellanic Cloud (LMC).  
LMC-2 is comparable to the D03 shell in linear size, but its X-ray 
luminosity is an order of magnitude lower \citep{Petal01}.  
In this case, the super-star cluster in the D03 supergiant shell 
must have played a significant role in the production of hot gas.  
Our echelle observations of this supergiant shell show larger 
velocity widths in the bright \ion{H}{2} region on the shell 
rim than within the supergiant shell.
Unfortunately, the spatial resolution of our echelle observations 
is much lower than that of the WFPC2 so the supergiant shell is 
not adequately resolved and only an upper limit of 50 km~s$^{-1}$ 
can be placed on its expansion velocity.

\subsection{Regions without Active Star Formation}

Three of the diffuse X-ray emission regions we selected, D06, D08, 
and D11, contain no obvious star formation activity, as indicated 
by the H$\alpha$ image.
The lack of active star formation in these regions is also
indicated by CO maps \citep{Letal01} and mid-IR (12--18 $\mu$m)
images \citep{ML99}.
No super-star clusters exist in D06 and D08, and no clusters are
cataloged in D11 at all.  
The origin of the hot gas in these regions without active star
formation is puzzling.  
It is clearly not powered by super-star clusters.
It is unlikely that this hot gas is transported from neighboring
star formation regions via breakouts, as these regions contain much
more thermal energy than the regions with star formation or 
super-star clusters.  

Not all massive stars are formed in clusters or super-clusters.
In nearby galaxies where individual massive stars are resolved,
it has been shown that giant \ion{H}{2} regions may contain only
loose OB associations without high concentrations, e.g., NGC\,604
in M33 \citep{Hetal96}, and that supergiant shells may be associated
with wide-spread, low-surface-density star formation, e.g., Shapley
Constellation III in LMC-4 \citep{S56}.
Both NGC\,604 and LMC-4 are diffuse X-ray sources with X-ray 
luminosities of 10$^{37}$--10$^{38}$ and $1.4\times10^{37}$ 
ergs~s$^{-1}$, respectively \citep{Yetal96,BDK94}.  
It is therefore possible that the hot gas in the Antennae regions 
without on-going active star formation was generated by field stars 
or low-mass clusters within the last $\sim$100 Myr

\section{Summary}

We have used $Chandra$ archival observations of NGC\,4038/4039, the 
Antennae galaxies, to investigate the distribution and physical 
properties of the hot interstellar gas.  
Eleven distinct diffuse X-ray emission regions are selected according 
to their underlying interstellar structures and star formation 
activity;  
five contain giant \ion{H}{2} regions, two are associated with 
circumnuclear starbursts, one is coincident with a supergiant shell, 
and three do not show active star formation. 
Their X-ray spectra are analyzed to determine the thermal energy 
content and cooling timescale.  

To investigate the stellar energy input in these diffuse X-ray 
emission regions, we have identified the underlying young star 
clusters and used the evolutionary stellar population synthesis 
code Starburst99 to model their photometric magnitudes and colors 
in order to assess their initial masses and ages.  
The cluster masses and ages are then used to determine the stellar 
wind and supernova energies injected into the ISM in each region.  

Comparisons between the thermal energy in the hot ISM and the 
expected stellar energy input show that in diffuse X-ray emission 
regions associated with active star formation, such
as giant \ion{H}{2} regions, circumnuclear starbursts, and 
supergiant shells, the underlying young star clusters play
significant to dominant roles in heating the ISM.
Super-star clusters with masses above $1\times10^5$ M$_\odot$ 
contribute to the production of hot ISM, but the yield of hot gas
is not linearly proportional to the cluster mass.  

There are regions without active star formation where 
the hot gas cannot have been produced by young star clusters.  
We suggest that the hot gas in these regions without on-going 
active star formation may have been produced by field stars or 
low-mass clusters.

\acknowledgments

We thank G.\ Bruzual and S.\ Charlot for providing their evolutionary 
stellar population synthesis code and D. Garnett for finding 
obscure references of the Antennae's abundances.
M.\ A.\ G.\ and R.\ A.\ G.\ are partially supported by the $Chandra$ 
grants SAO AR3-40001X and SAO GO3-4023X.  

%\newpage
\appendix

\section{Choosing a Spectral Model}

The physical properties of an X-ray emitting plasma can be determined
from the best model fit to the observed spectrum.
To improve the spectral fits, models for diffuse X-ray sources frequently 
include multiple thermal plasma components and a power law component,
the latter of which is argued to represent the emission from unresolved 
point sources.
These complex models almost always lead to better fits with 
unrealistically low abundances, for example, NGC 253 \citep{Setal02} and 
NGC 4631 \citep{Wetal01}. 
It is thus worthwhile to examine the effects and necessity of these 
spectral components.

We use the spectrum of region D06 to illustrate how models affect the 
results of spectral fitting.
We have fit the spectrum with the following models:
(1) single thermal plasma component, (2) two thermal plasma components, 
(3) single thermal plasma component plus a power law, and
(4) two thermal plasma components plus a power law.  
For each of these models, we try spectral fits with floating and fixed
plasma abundances, and with floating and fixed absorption column density.
The MEKAL model \citep{KM93,LOG95} is used for the thermal plasma 
emission, the absorption cross sections are from \citet{BM92}
and the solar abundances are adopted for the foreground absorption.
The additional absorption due to the build-up on the ACIS-S is also
modeled and taken into account in the spectral fits.

When we make spectral fits with floating absorption column 
density, the best-fits all converge to a column density much lower 
than the foreground Galactic absorption, which is unphysical.  
We have therefore adopted the Galactic \ion{H}{1} column density
of $3.9\times10^{20}$ cm$^{-2}$ \citep{DL90} as the absorption column 
density.
The results discussed below are based on models using this fixed 
absorption column density.

For a single thermal plasma component model, the best spectral fit 
with floating abundances gives a plasma temperature of $kT$ = 0.55 keV 
but an unphysically low abundance of 0.14 Z$_\odot$.
The best fit with an abundance fixed at the solar value gives a
similar temperature, $kT$ = 0.55 keV, but the model underestimates 
emission at 0.5--0.6 keV and $\sim$1.5 keV. 
See the top two panels of Figure~6 for a comparison between 
spectral fits with abundances of 0.14 Z$_\odot$ and 1 Z$_\odot$.
The 0.14 Z$_\odot$ model fits the observed spectrum better because 
the metal lines between 0.6 keV and 1 keV are suppressed so that
the relative contribution from the bremsstrahlung emission can be
raised to produce adequate emission below 0.6 keV.

For a model with two thermal plasma components each with independent 
floating abundances, the best fit converges to one component with 
$kT_1$ = 0.37~keV and abundances of 1.7 Z$_\odot$, and another component
with higher temperature $kT_2$ = 0.71~keV and abundances of 0.7 
Z$_\odot$.  
The middle-left panel of Figure~6 shows an example of a spectral fit 
with two thermal plasma components of different abundances.  The spectra 
of the two plasma components are individually plotted to show their 
contributions to the total emission (middle-left panel of Figure~6).
The lower temperature component prevails in the low energy band, 
while the higher temperature component dominates the $>$ 1 keV 
band.  

For a model with two thermal plasma components of identical floating
abundances, the best fit has temperatures $kT_1 =$ 0.38 keV and 
$kT_2$ = 0.68 keV, but unphysically low abundances $Z$ = 0.19 $Z_\odot$. 
As in the above case, the higher temperature component dominates 
the $>$ 1 keV band (middle-right panel of Figure~6).  

For a model with two thermal plasma components of solar abundances, 
the best fit converges to one component with $kT_1$ = 0.23~keV and a 
weaker component with $kT_2$ = 0.61~keV.
The bottom-left panel of Figure~6 shows the spectra of these individual
components.
The main component accounts for the overall spectral shape, and
the secondary component produces more emission below 0.6 keV.

Adding a power law component to the single thermal plasma component model, 
the best fit has a temperature, $kT$ $\sim$ 0.5 keV, similar to that of 
the model without a power law component, but the abundance is significantly 
higher, $Z$ = 1.2 $Z_\odot$.
Fixing the abundance at the solar value, the best spectral fit also gives
a temperature, $kT \sim$ 0.52 keV, similar to that of the model without a 
power law component. 
The contributions of the thermal plasma and power law components are
illustrated in the bottom-right panel of Figure~6.
It is evident that the power law component facilitates the fit by 
contributing to the soft X-ray emission.

Finally, using a model incorporating a power law component and two thermal 
components, the best fit shows a main thermal plasma component and a power
law component similar to those for the above model with a single thermal 
plasma component and a power law component, and a secondary thermal 
component which makes a negligible contribution to the total emission.

It becomes clear from these results that the spectral ranges below and
above 0.6 keV cannot be simultaneously fit well by a model with a 
single thermal plasma component at the solar abundance.
The spectral fit below 0.6 keV can be improved by drastically lowering 
the abundance to suppress the metal lines above 0.6 keV, adding another
thermal plasma component at a low temperature to boost the X-ray
emission below 0.6 keV, or adding a power law component that rises toward 
the low energies.

Are these ad hoc fixes artificial?
Abundances as low as 0.1 Z$_\odot$ are commonly found in dwarf galaxies
such as the Small Magellanic Cloud, but not in gas-rich spirals such
as the Antennae galaxies.
Spectrophotometric observations of 18 \ion{H}{2} regions in the Antennae
show an [\ion{O}{3}] $\lambda$5007 line comparable to or weaker than the 
H$\beta$ line \citep{Retal70}, which is consistent with low electron 
temperatures and high abundances.
Moreover, the abundances of a dwarf galaxy in the tidal tail drawn
from the outer parts of the Antennae are comparable to the LMC 
abundance \citep{MDL92}, further supporting that the abundances
of the inner parts of the Antennae should be solar or higher.
Therefore, the X-ray spectral fits that require metallicities below 
0.3 Z$_\odot$ can be ruled out.

The addition of a power law component is also not justified because
the diffuse X-ray emission from the Antennae galaxies is associated 
with star formation regions and is not expected to include a significant
contribution from faint, unresolved point sources of X-ray binaries.
For example, $ROSAT$ observations of the LMC (at 50 kpc) show that 
X-ray binaries brighter than 10$^{34}$ ergs~s$^{-1}$ are rare in 
star forming regions and no power law component is needed in the 
analysis of the diffuse X-ray emission \citep{Chu98,CS98,Petal01}.

As it is not even clear whether the ACIS-S spectra in the soft energy range 
are accurately calibrated \citep{Setal02}, it is premature to include
multiple components and adhere to the absolute least-$\chi^2$ for 
the best model fit.
Therefore, we have chosen to analyze the spectra with only one single 
thermal plasma component at a fixed abundance, $Z = Z_\odot$.

It is possible that the abundance is higher or lower than the solar value.
As the X-ray emissivity scales directly with the abundance $Z$,
the normalization factor of the spectral fit, or the volume emission
measure, will scale with $Z^{-1}$.
Consequently, the rms density, mass, and thermal energy of the 
X-ray-emitting plasma all scale with $Z^{-1/2}$.
Thus the uncertainties in the mass and thermal energy are not as large as 
the uncertainty in the abundance itself.

\section{Comparisons among Different Models}

Following the referee's suggestion, we compare the goodness-of-fit among
different models and show how these different models affect the thermal
energy of the hot gas. 
First we carry out single-component MEKAL model fits with fixed solar 
abundances and with floating (free-varying) abundances for all 11
diffuse emission regions. 
The results are presented in Table 6.
By allowing the abundances to vary freely and converge at very low 
abundances, $\sim 0.1~Z_\odot$, the goodness-of-fit can be 
significantly improved.
Compared to the best-fit models with fixed solar abundance, the 
best-fit models with free-varying abundances have similar temperatures
but the thermal energies are 2--4 times higher.
The thermal energies are raised because higher densities are needed to
compensate for the low abundances in order to produce the observed emission,
and higher densities lead to larger thermal energies.
The density and thermal energy roughly scale with $Z^{-1/2}$, as explained 
at the end of Appendix A, 

As we discussed in Appendix A, abundances as low as 0.1 $Z_\odot$ can be
excluded for the Antennae galaxies because even the outer parts of the
Antennae have higher abundances \citep{MDL92}.
Thus, we next consider one-component and 2-component MEKAL models with
solar abundances for all 11 diffuse X-ray sources.
The results are presented in Table 7.
The goodness-of-fit is improved in the 2-component model fits.
The two temperatures in the 2-component model fits bracket the 
temperatures of the corresponding 1-component model fits.
In some of the cases, the second component has unrealistically high
temperatures, which may be caused by low S/N in the data or simply
deficiencies of the MEKAL model.
When the two temperatures of the 2-component model are reasonable, 
the resulting thermal energy is higher, but not more than a factor 
of 2 higher, than that of the corresponding 1-component model.
This difference does not affect the conclusions of our comparison
between stellar energy input and the thermal energy in \S5.4.

Finally, we compare spectral fits made with the APEC (Astrophysical 
Plasma Emission Code) model \citep{Setal01} and with the MEKAL model. 
Two temperature components are used with solar abundances.
The results are presented in Table 8.
The APEC spectral fits are generally consistent with the MEKAL 
spectral fits, and the resultant thermal energies are similar,
although the APEC models fit the spectra better (smaller reduced
$\chi^2$), and give more reasonable temperatures even when MEKAL 
models produce unrealistically high temperatures.
Note that the APEC models still produce very high temperatures for
D05, D09, and D11, which might be due to the low S/N in the spectra
of D05 and D09 and uncertain origins of the hot gas in D11.

%%%%%%%%%%%%%%%%%%%%%%%%%%%%%%%%%%%%%%%%%%%%%%%%%%%%%%%%%%%%%%%%%%%%%%%%

\newpage

\begin{figure}
\figurenum{1}
\centerline{\plotone{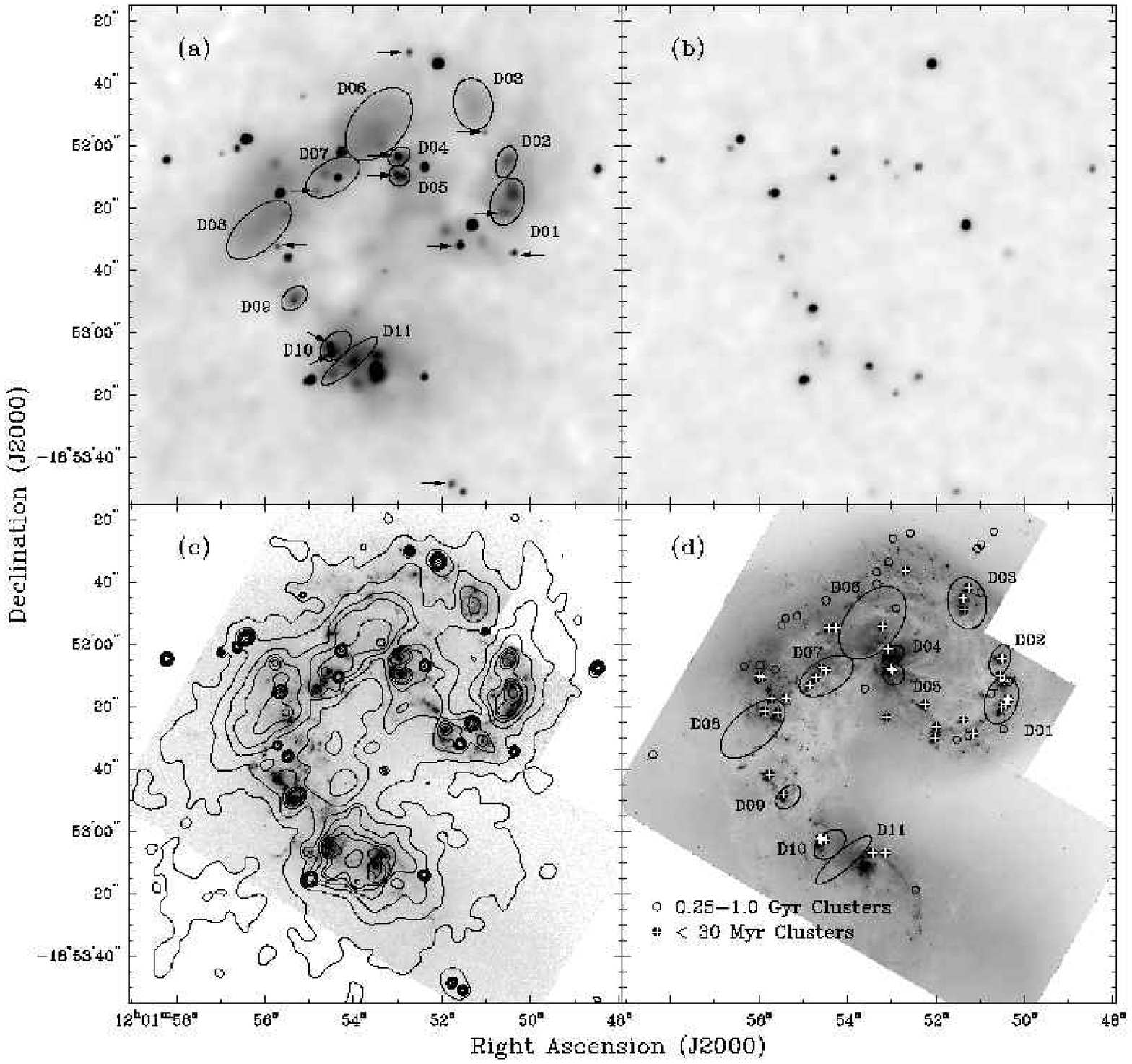}}
\caption{
The top two panels contain adaptively smoothed X-ray images of 
the Antennae galaxies in the 0.3--2.0 keV energy band (top left) 
and 2.5--6.0 keV energy band (top right).  See the text for details
of this merged X-ray image.
The pixel size is 0\farcs492 and the maximum smoothing scale is 2\arcsec.  
The locations of the 11 elliptical regions that have been selected for 
study are shown in the soft energy image.  
The positions of soft point sources that are possibly SNR candidates 
are marked by arrows in the soft band image.
The bottom left panel presents an \emph{HST} WFPC2 H$\alpha$ image
(\emph{F658N}) overlaid by X-ray contours for the 0.3--2.0 keV band.
The contour levels are 0.2, 0.45, 0.8, 1.2, 1.8, 2.5, 4.5, 10, 40,
and 150 counts pixel$^{-1}$.
The bottom right panel displays an \emph{HST} WFPC2 $V$-band 
(\emph{F555W})
image with the locations of clusters and the 11 diffuse X-ray emission 
regions marked.
}
\end{figure}

%\clearpage

\begin{figure}
\figurenum{2}
\centerline{\epsscale{0.90}\plotone{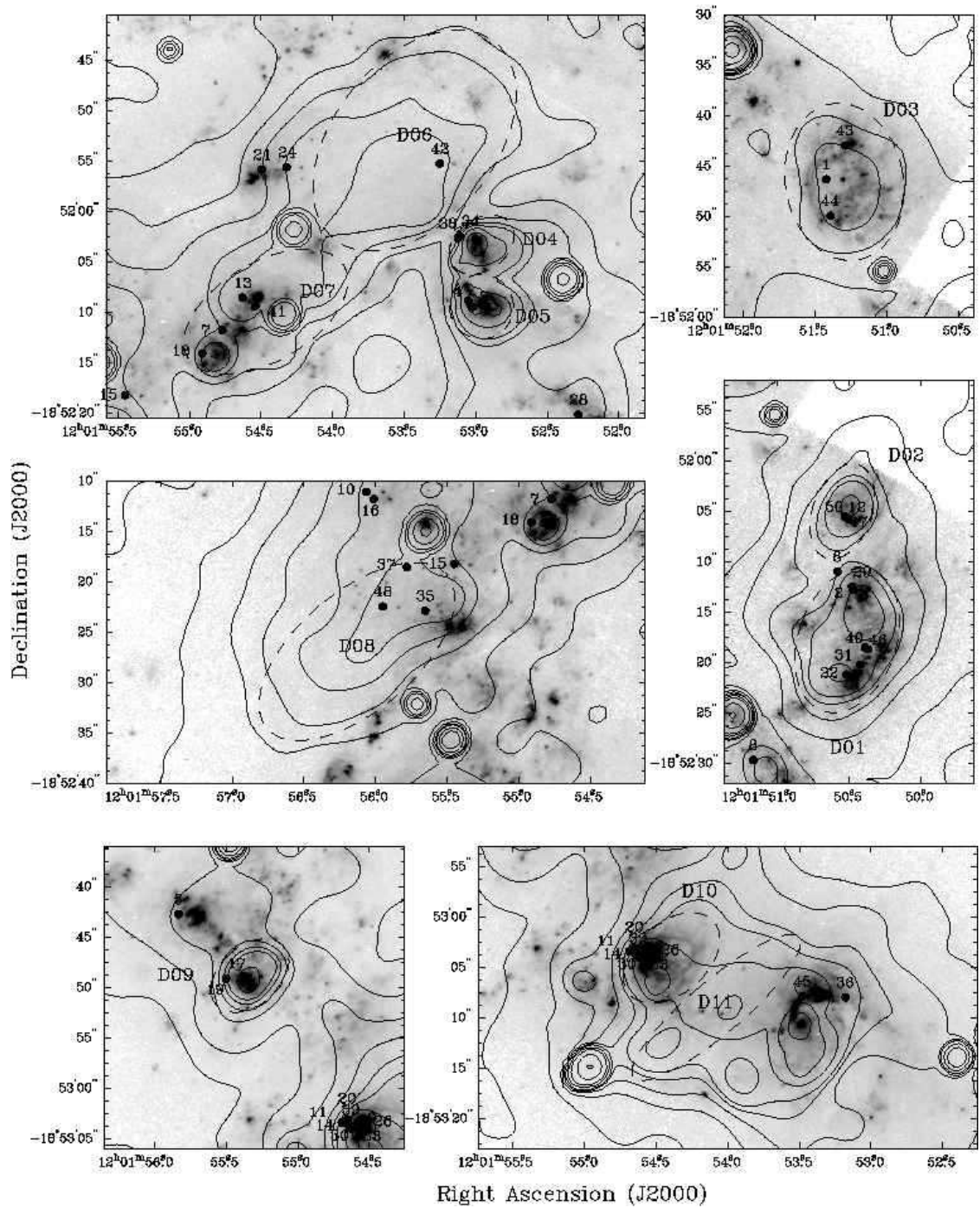}\epsscale{1.0}}
\caption{
H$\alpha$ images with soft X-ray contours (solid lines) of the 
11 diffuse X-ray emission regions that have been chosen for study 
(black dashed lines).  
The contour levels are 0.16, 0.35, 0.65, 1.0, 1.45, 1.95, 3.5, 7.0, 
30, and 120 counts pixel$^{-1}$, and the pixel size is 0\farcs492.
See Table 1 for regions of point sources that have been excised from
our analysis of diffuse X-ray emission.
The black dots and numbers represent the location and catalog number 
of the young star clusters as listed in Table 1 of \citet{Whitmore99}
and in Table 4 of this paper.
}
%\label{}
\end{figure}

%\clearpage

\begin{figure}
\figurenum{3}
\centerline{\plottwo{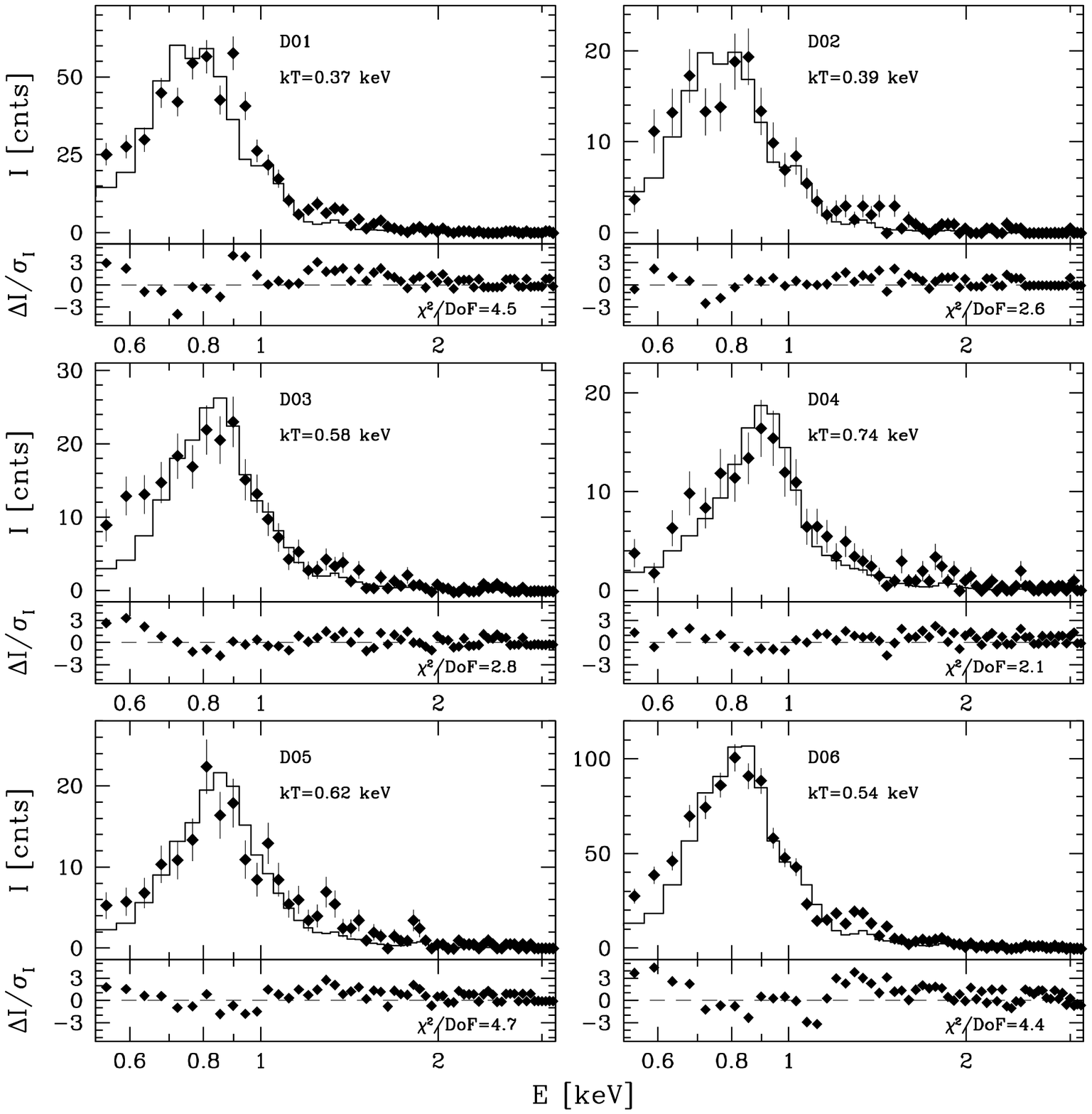}{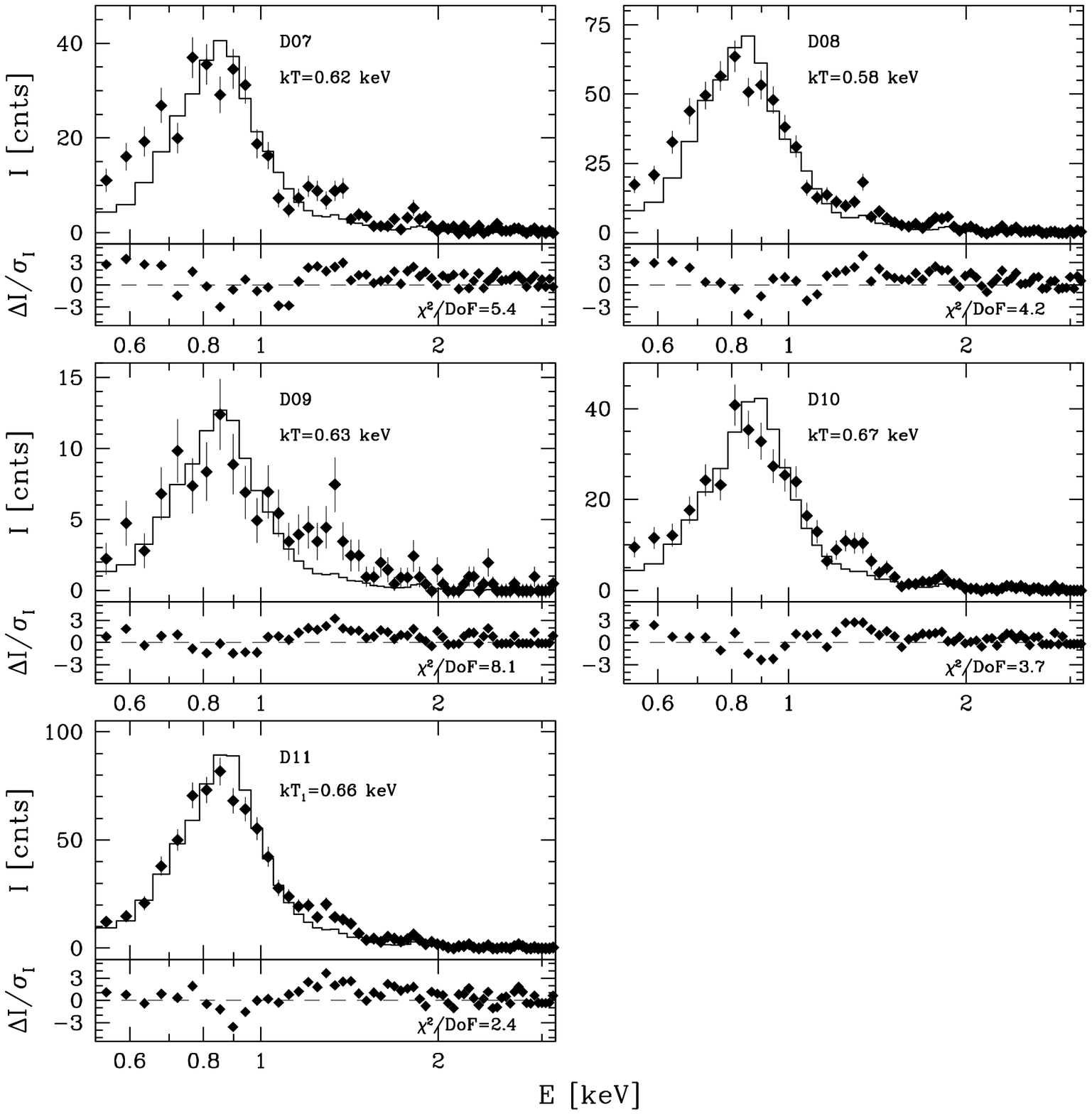}}
\caption{
Spectra and best-fit models of the 11 regions with diffuse X-ray 
emission.  The regions are listed in Table 1 and described in \S3.
The adopted spectral model consists of a single-temperature MEKAL
model with a fixed solar abundance.
The spectra and best fits are shown in the upper panels, and the 
residuals of the fits in units of $\sigma$ for each energy bin
are shown in the lower panels.  
}  
\label{fig_spec1}
\end{figure}

\begin{figure}
\figurenum{4}
\centerline{\plotone{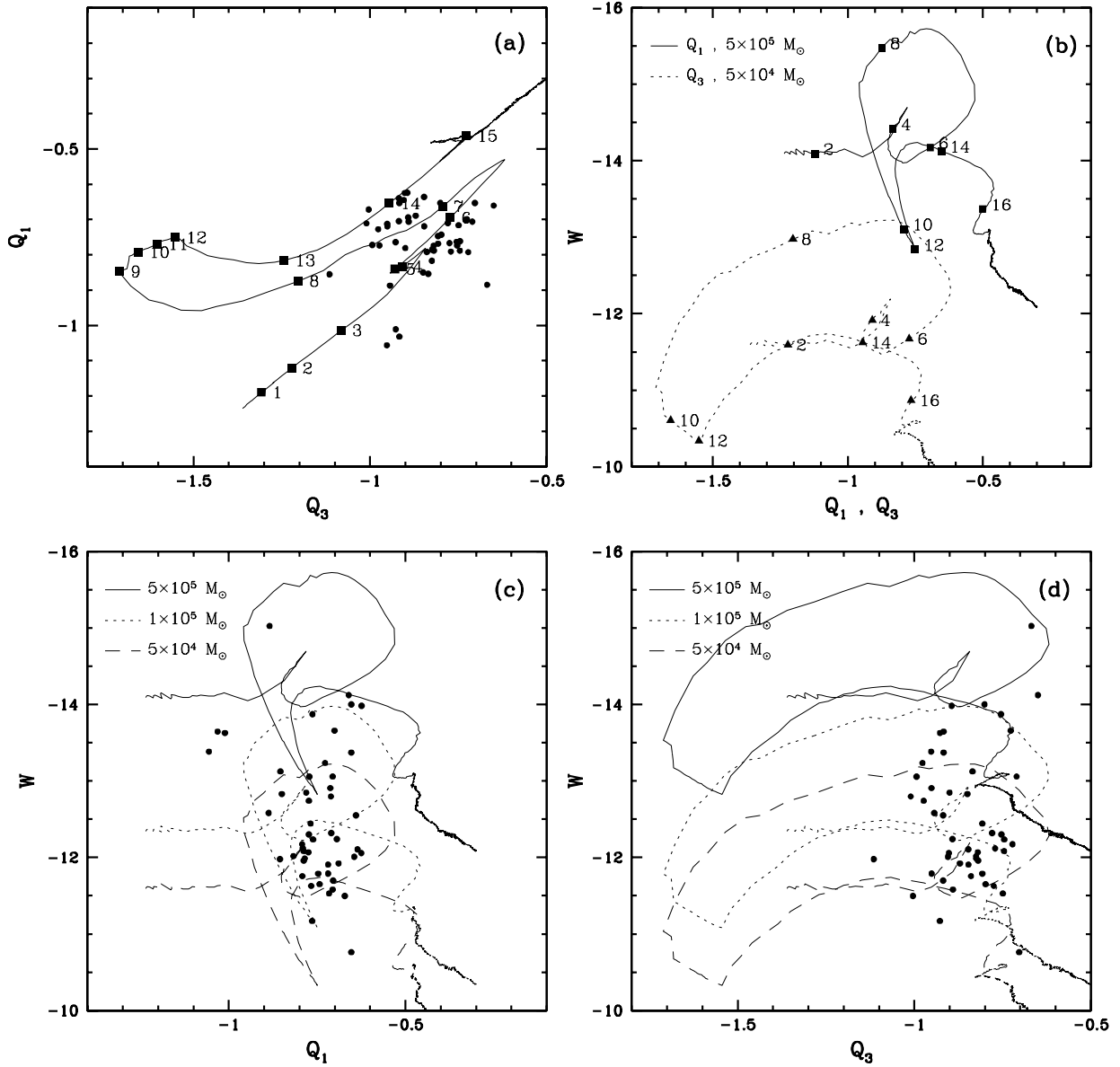}}
\caption{(a) Reddening-free color-color diagram for $Q_1$ versus
$Q_3$.  The 50 young star clusters from Table 1 of \citet{Whitmore99} 
are plotted as solid circles.
The evolutionary track produced by Starburst99 for a cluster with 
solar abundance is plotted with ages marked every Myr as solid 
squares.
(b) Reddening-free color-magnitude diagrams for $W$ versus $Q_1$
 and $W$ versus $Q_3$.
The evolutionary track for a cluster of initial masses of 
$5\times10^5$ M$_\odot$ in the $W$--$Q_1$ diagram is plotted 
in solid curve, and that for a $5\times10^4$ M$_\odot$ cluster 
is plotted in the $W$--$Q_3$ diagram in dashed curve. 
(Different masses are used in these two evolutionary tracks to
offset the curves vertically for clarity.)
The cluster ages are marked along the  evolutionary tracks 
in 2 Myr intervals.
(c) Reddening-free color-magnitude diagram for $W$ versus $Q_1$.
 The 50 young star clusters from Table 1 of \citet{Whitmore99}
 are plotted as solid circles.
 Evolutionary tracks for clusters of initial masses of
 $5\times10^4$, $1\times10^5$, and $5\times10^5$ M$_\odot$
 are plotted .
(d) same as (c) for $W$ versus $Q_3$.
}
%\label{fig_spec2}
\end{figure}

\begin{figure}
\figurenum{5}
\centerline{\epsscale{0.90}\plotone{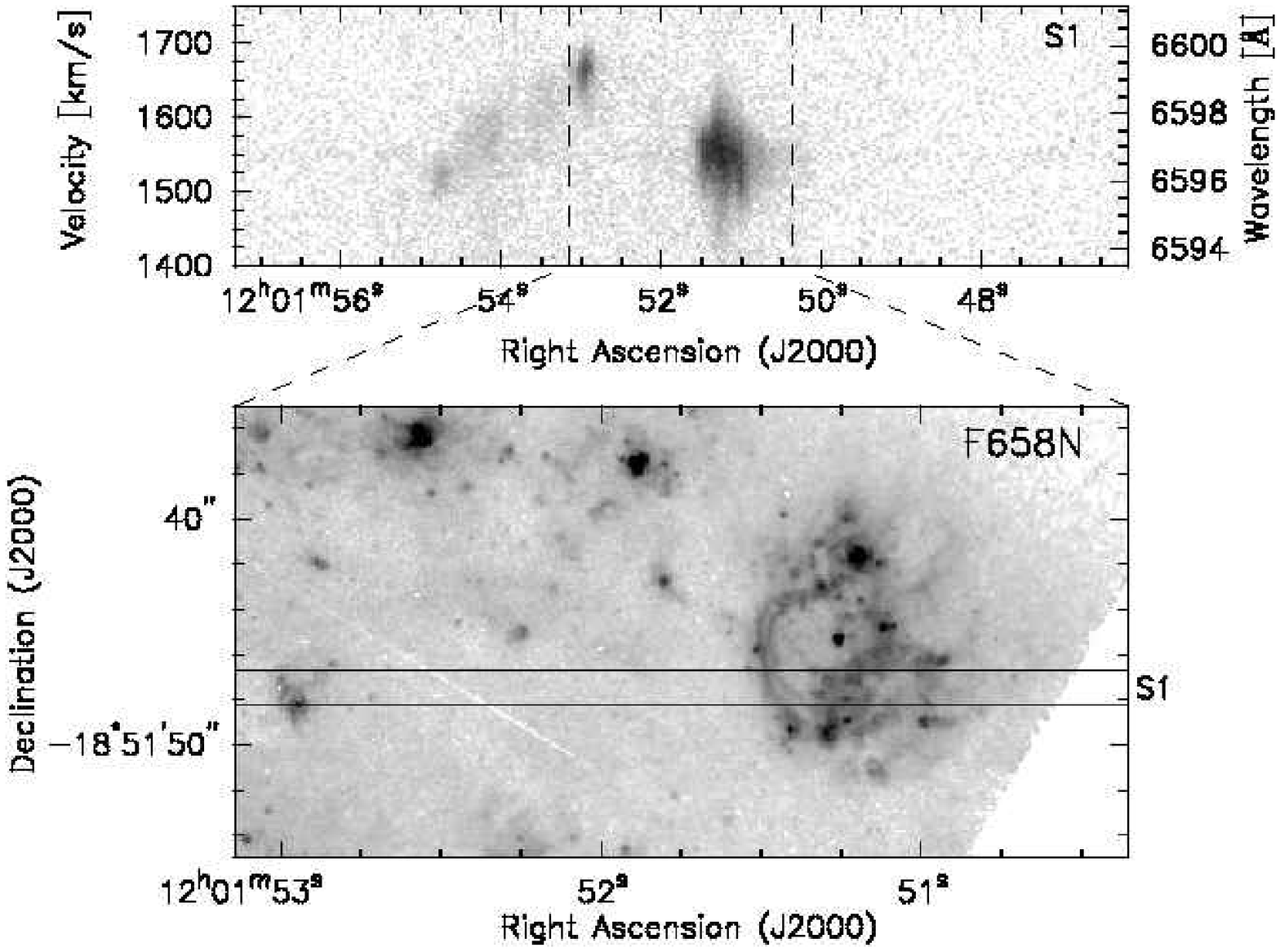}\epsscale{1.0}}
\centerline{\epsscale{0.90}\plotone{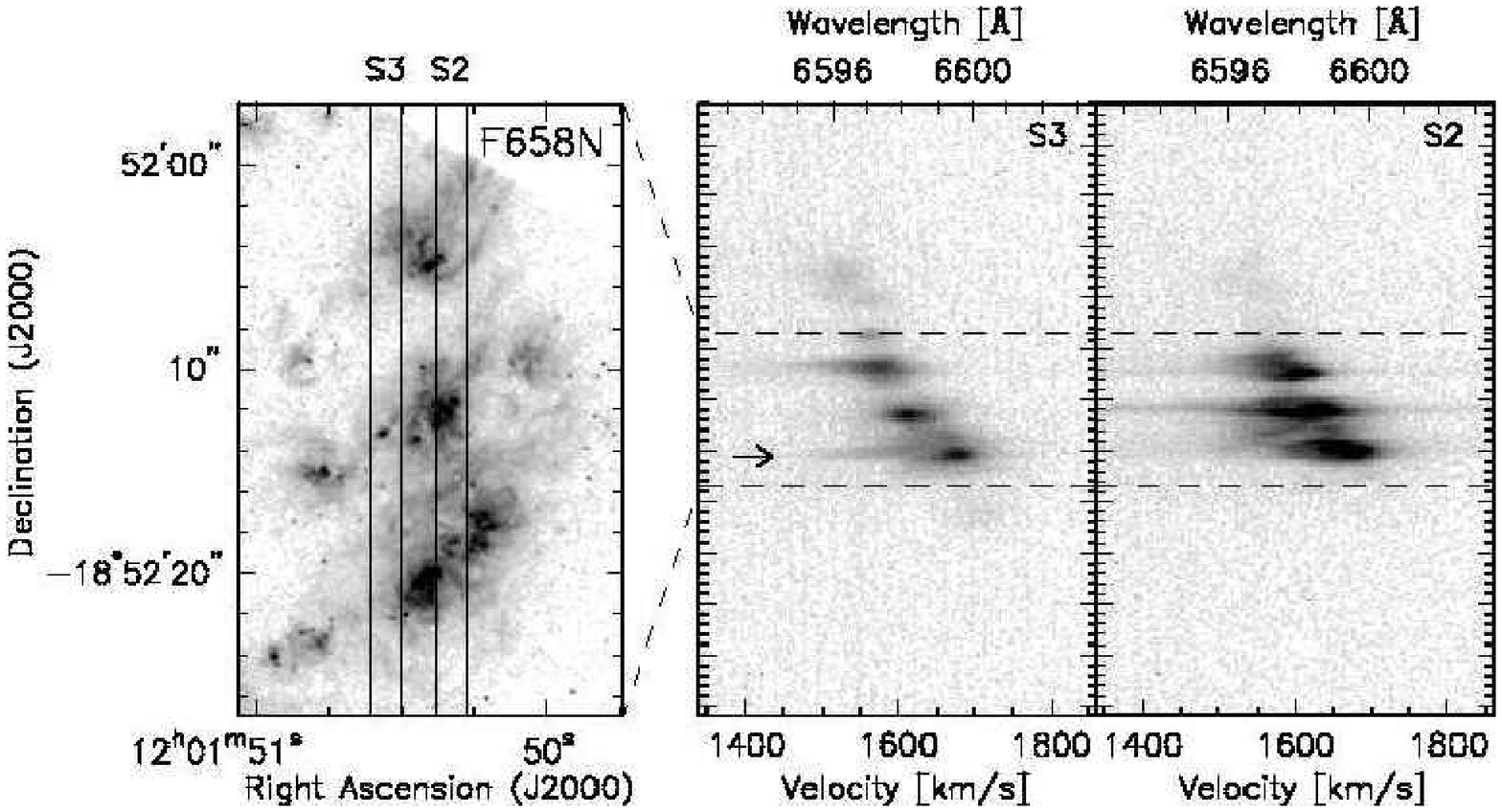}\epsscale{1.0}}
\caption{Echellograms of the H$\alpha$ line region at slit 
positions S1, S2, and S3 along with \emph{HST} WFPC2 \emph{F658N} 
images showing the location of the slit positions. 
Slit position S1 covers the supergiant shell in D03.
Slit positions S2 and S3 sample the giant \ion{H}{2} regions
and their vicinities in D01 and D02.
The arrow in the echellogram for the slit position S3 marks the 
high-velocity gas associated with a discrete soft X-ray 
source, suggesting a SNR candidate.}
\label{fig_spec2}
\end{figure}

\begin{figure}
\figurenum{6}
\centerline{\plotone{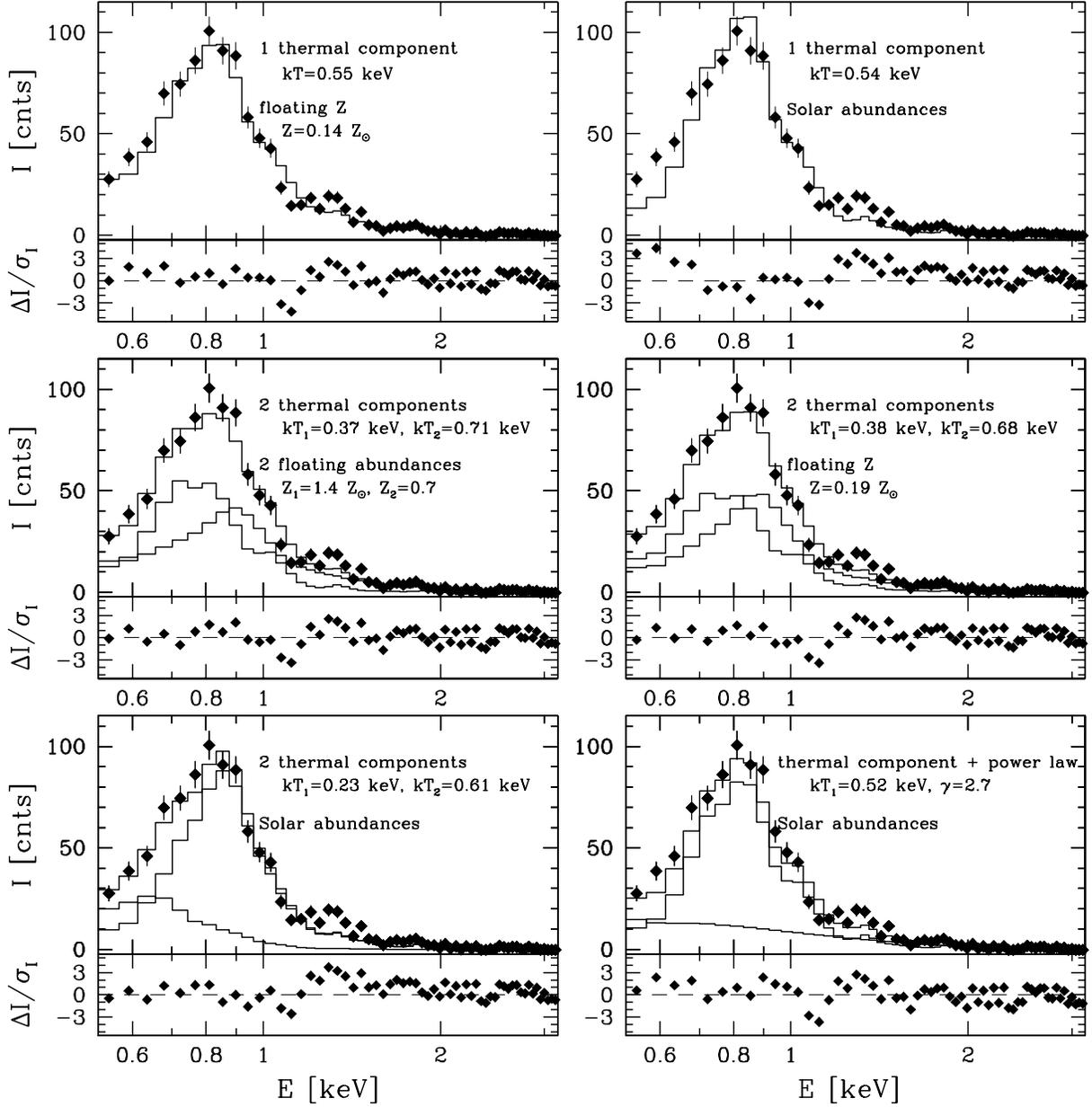}}
\caption{
Spectral fits for region D06 using different models.  
As in Figure~3, the top panel shows the spectra and spectral fits, and 
the bottom panel shows the residuals in units of $\sigma$ for each
energy bin.  
The model and the best-fit parameters are labeled in each panel.
For spectral fits with more than one component, the individual 
components and their sum are all shown.  
}
\label{fig_spec0}
\end{figure}

%\clearpage

\begin{deluxetable}{lcccrrrl}
\tablewidth{0pt}
\tablecaption{Regions of Diffuse X-ray Sources in the Antennae Galaxies}
\tablehead{
\multicolumn{1}{c}{Region} & 
\multicolumn{1}{c}{R.A.} & 
\multicolumn{1}{c}{Dec} & 
\multicolumn{1}{c}{Size} & 
\multicolumn{1}{c}{P.A.} & 
\multicolumn{1}{c}{$V$\,\tablenotemark{a}} & 
\multicolumn{1}{c}{Counts} & 
\multicolumn{1}{c}{Apparent} 
\\
\multicolumn{1}{c}{} & 
\multicolumn{2}{c}{(J2000)} & 
\multicolumn{1}{c}{} & 
\multicolumn{1}{c}{($^\circ$)} & 
\multicolumn{1}{c}{(pc$^3$)} & 
\multicolumn{1}{c}{} & 
\multicolumn{1}{c}{Association}

}
\startdata
~D01 &    12 01 50.51 & $-$18 52 17.9~ &  14\farcs8$\times$~9\farcs8~ &  
  160~ & 6.0$\times$10$^8$ & 1150~ & Giant \ion{H}{2} region \\
~D02 &    12 01 50.55 & $-$18 52 05.1~ &  ~9\farcs8$\times$~5\farcs9~ &  
  160~ & 1.4$\times$10$^8$ &  375~ & Giant \ion{H}{2} region \\
~D03 &    12 01 51.29 & $-$18 51 46.6~ &  15\farcs7$\times$11\farcs8~ &  
   10~ & 9.2$\times$10$^8$ &  470~ & Supergiant shell \\
~D04\tablenotemark{b}
     &    12 01 52.96 & $-$18 52 03.3~ &  ~7\farcs9$\times$~5\farcs9~ &  
  125~ & 1.1$\times$10$^8$ &  355~ & Circumnuclear \ion{H}{2}  \\
~D05\tablenotemark{c} &
          12 01 52.96 & $-$18 52 09.5~ &  ~6\farcs9$\times$~6\farcs9~ & 
  \nodata & 1.3$\times$10$^8$ &  410~ & Circumnuclear \ion{H}{2} \\
~D06 &    12 01 53.43 & $-$18 51 53.0~ &  25\farcs3$\times$16\farcs0~ &  
  140~ & 2.7$\times$10$^9$ & 1950~ & \nodata \\
~D07\tablenotemark{d} &
          12 01 54.47 & $-$18 52 10.0~ &  18\farcs1$\times$~9\farcs8~ &  
  120~ & 7.0$\times$10$^8$ &  790~ & Giant \ion{H}{2} region \\
~D08 &    12 01 56.13 & $-$18 52 27.0~ &  23\farcs6$\times$11\farcs8~ &  
  130~ & 1.4$\times$10$^9$ & 1320~ &  \nodata \\
~D09\tablenotemark{e} &
          12 01 55.33 & $-$18 52 48.9~ &  ~8\farcs9$\times$~5\farcs9~ &  
  130~ & 1.2$\times$10$^8$ &  275~ & Giant \ion{H}{2} region \\
~D10\tablenotemark{f} &
          12 01 54.40 & $-$18 53 04.1~ &  11\farcs8$\times$~7\farcs9~ &  
  130~ & 2.9$\times$10$^8$ &  800~ & Giant \ion{H}{2} region \\
~D11 &    12 01 54.08 & $-$18 53 09.0~ &  21\farcs6$\times$~5\farcs9~ &    
  130~ & 3.2$\times$10$^8$ & 1620~ & \nodata \\
\enddata
\tablenotetext{a}{The volume is calculated assuming an ellipsoidal geometry
 with the third axis equal to the minor axis of the elliptical region.}
\tablenotetext{b}{A circular region centered at 12$^{\rm h}$01$^{\rm 
m}$53\rlap{.}{$^{\rm s}$}01, $-$18$^\circ$52$'$02\farcs8 with r=1\farcs2\ is excised.}
\tablenotetext{c}{A circular region centered at 12$^{\rm h}$01$^{\rm 
m}$53\rlap{.}{$^{\rm s}$}01, $-$18$^\circ$52$'$09\farcs5 with r=1\farcs2\ is excised.}
\tablenotetext{d}{Circular regions centered at 
12$^{\rm h}$01$^{\rm m}$54\rlap{.}{$^{\rm s}$}83, $-$18$^\circ$52$'$14\farcs2 with r=1\farcs2, 
12$^{\rm h}$01$^{\rm m}$54\rlap{.}{$^{\rm s}$}67, $-$18$^\circ$52$'$9\farcs0  with r=1\farcs2, and 
12$^{\rm h}$01$^{\rm m}$54\rlap{.}{$^{\rm s}$}36, $-$18$^\circ$52$'$10\farcs4 with r=1\farcs7\ 
are excised.}
\tablenotetext{e}{A circular region centered at 12$^{\rm h}$01$^{\rm 
m}$55\rlap{.}{$^{\rm s}$}19, $-$18$^\circ$52$'$47\farcs4 with r=1\farcs7\ is excised.}
\tablenotetext{f}{Circular regions centered at 
12$^{\rm h}$01$^{\rm m}$54\rlap{.}{$^{\rm s}$}54, $-$18$^\circ$53$'$03\farcs9 and 
12$^{\rm h}$01$^{\rm m}$54\rlap{.}{$^{\rm s}$}52, $-$18$^\circ$53$'$07\farcs1 
  with r=1\farcs5\ are excised.}
\end{deluxetable}

\begin{deluxetable}{lccclcr}
\tablewidth{0pt}
\tablecaption{Results of X-ray Spectral Fits}
\tablehead{
\multicolumn{1}{c}{Region} & 
\multicolumn{1}{c}{$T$} & 
\multicolumn{1}{c}{$kT$} & 
\multicolumn{1}{c}{$A$} & 
\multicolumn{1}{c}{rms $N_{\rm e}$} & 
\multicolumn{1}{c}{$f^{\rm obs}_{\rm X}$ (0.5--2.0 keV)} & 
\multicolumn{1}{c}{$L_{\rm X}$ (0.5--2.0 keV)} \\
\multicolumn{1}{c}{} & 
\multicolumn{1}{c}{(10$^6$ K)} & 
\multicolumn{1}{c}{(keV)} & 
\multicolumn{1}{c}{(cm$^{-5}$)} & 
\multicolumn{1}{c}{(cm$^{-3}$)} & 
\multicolumn{1}{c}{(ergs\,cm$^{-2}$\,s$^{-1}$)} & 
\multicolumn{1}{c}{(10$^{38}$ ergs\,s$^{-1}$)}
}

\startdata
~D01 & 4.3$\pm$0.5 & 0.37$\pm$0.04 & 9.9$\times$10$^{-6}$ & ~~0.05  & 2.0$\times$10$^{-14}$ & 13.0~~~~~~~~ \\ 
~D02 & 4.5$\pm$1.0 & 0.39$\pm$0.09 & 3.1$\times$10$^{-6}$ & ~~0.06  & 6.5$\times$10$^{-15}$ &  4.1~~~~~~~~ \\ 
~D03 & 6.7$\pm$1.0 & 0.58$\pm$0.09 & 3.5$\times$10$^{-6}$ & ~~0.02  & 8.3$\times$10$^{-15}$ &  5.1~~~~~~~~ \\ 
~D04 & 8.6$\pm$0.9 & 0.74$\pm$0.08 & 2.7$\times$10$^{-6}$ & ~~0.06  & 6.3$\times$10$^{-15}$ &  3.8~~~~~~~~ \\ 
~D05 & 7.2$\pm$0.9 & 0.62$\pm$0.08 & 2.9$\times$10$^{-6}$ & ~~0.06  & 6.9$\times$10$^{-15}$ &  4.2~~~~~~~~ \\ 
~D06 & 6.3$\pm$0.8 & 0.54$\pm$0.07 & 1.4$\times$10$^{-5}$ & ~~0.03  & 3.3$\times$10$^{-14}$ & 20.0~~~~~~~~ \\ 
~D07 & 7.2$\pm$0.5 & 0.62$\pm$0.04 & 5.2$\times$10$^{-6}$ & ~~0.03  & 1.3$\times$10$^{-14}$ &  7.7~~~~~~~~ \\ 
~D08 & 6.7$\pm$0.6 & 0.58$\pm$0.05 & 9.2$\times$10$^{-6}$ & ~~0.03  & 2.2$\times$10$^{-14}$ & 13.0~~~~~~~~ \\ 
~D09 & 7.3$\pm$1.2 & 0.63$\pm$0.10 & 1.7$\times$10$^{-6}$ & ~~0.04  & 4.1$\times$10$^{-15}$ &  2.5~~~~~~~~ \\ 
~D10 & 7.8$\pm$0.6 & 0.67$\pm$0.05 & 5.7$\times$10$^{-6}$ & ~~0.05  & 1.3$\times$10$^{-14}$ &  8.2~~~~~~~~ \\ 
~D11 & 7.7$\pm$0.4 & 0.66$\pm$0.03 & 1.2$\times$10$^{-5}$ & ~~0.08  & 2.8$\times$10$^{-14}$ & 17.0~~~~~~~~ \\ 
\enddata
\end{deluxetable}

\begin{deluxetable}{lccccr}
%\tablecolumns{12}
\tablewidth{0pc}
\tablecaption{Energies and Cooling Timescales of Diffuse X-ray Emission 
Regions}
\tablehead{
\colhead{Region} & 
\colhead{Young Star} & 
\colhead{$E_{\rm th}$} & 
\colhead{$E_\star$ Input\tablenotemark{b}} &
\colhead{$E_\star$ Input\tablenotemark{c}} &
\colhead{$t_{\rm cool}$} \\
\colhead{} &
\colhead{Clusters\tablenotemark{a}} &
\colhead{~~~~(ergs)~~~~} &
\colhead{~~~~(ergs)~~~~} &
\colhead{~~~~(ergs)~~~~} &
\colhead{(Myr)}
}

\startdata
~D01 &       {\bf 2},22,29,31,40,46 & 1.5$\times$10$^{54}$ 
  & 1.4$\times$10$^{54}$ & 1.4$\times$10$^{54}$ &  65~~   \\ 
~D02 &        {\bf 12},47,50 & 4.3$\times$10$^{53}$ 
  & 1.0$\times$10$^{54}$ & 8.1$\times$10$^{53}$ &  55~~   \\ 
~D03 &        {\bf 1},43,44 & 1.7$\times$10$^{54}$ & 
  5.6$\times$10$^{54}$ & 6.7$\times$10$^{54}$ & 240~~     \\ 
~D04 &        {\bf 34},{\bf 38} & 6.6$\times$10$^{53}$ 
  & 4.0$\times$10$^{54}$ & 3.4$\times$10$^{54}$ &  105~~   \\ 
~D05 &      {\bf 3},{\bf 4} & 6.2$\times$10$^{53}$ & 
  3.8$\times$10$^{54}$ & 3.3$\times$10$^{54}$ &  90~~     \\ 
~D06 &          42 & 5.5$\times$10$^{54}$ & 1.0$\times$10$^{53}$
     & 1.1$\times$10$^{53}$ & 150~~     \\ 
~D07 &    {\bf 7},13,{\bf 18},41 & 1.9$\times$10$^{54}$ 
     & 1.8$\times$10$^{54}$ & 1.3$\times$10$^{54}$ & 175~~  \\ 
~D08 &         35,48 & 3.4$\times$10$^{54}$ & 9.4$\times$10$^{53}$ 
   & 1.1$\times$10$^{54}$ & 140~~ \\ 
~D09 &    {\bf 17},19 & 4.5$\times$10$^{53}$ & 9.5$\times$10$^{53}$ 
     & 7.8$\times$10$^{53}$ & 130~~  \\ 
~D10 & {\bf 9},{\bf 11},{\bf 14},{\bf 20},26,30,33,39
     & 1.4$\times$10$^{54}$ & 8.6$\times$10$^{54}$ & 5.1$\times$10$^{54}$
     &  115~~  \\ 
~D11 &  \nodata & 2.2$\times$10$^{54}$ & \nodata              
     & \nodata    &  70~~  \\ 
\enddata
\tablenotetext{a}{
Clusters from Table~1 of \citet{Whitmore99}, also listed in Table~4 
of this paper.  
Super-star clusters with masses $\ge 1\times10^5$ M$_\odot$ are 
highlighted in boldface.}
\tablenotetext{b}{Total stellar energy input, i.e., the sum of the 
stellar wind and supernova energies, estimated using the method 
outlined in \S5.2 and \S5.3.}  
\tablenotetext{c}{Total stellar energy input calculated from Starburst99.} 

%\tablecomments{
%(1) $\Lambda(T_e)=2.5\times10^{-23}$\\
%(2) All values calculated using $f=1.0$}

\end{deluxetable}

\begin{deluxetable}{rcrcrrrrrl}
\tablewidth{0pt}
\tablecaption{Stellar Wind and Supernova Energies of Young Star Clusters}
\tablehead{
\colhead{Cluster\tablenotemark{a}} & 
\colhead{$M_{\rm cluster}$} & 
\colhead{$t_{\rm cluster}$} & 
\colhead{$M_{\rm min}$} & 
\colhead{$N_{\rm SN}$} & 
\colhead{$E_{\rm SN}$\tablenotemark{b}} &
\colhead{$E_{\rm wind}$\tablenotemark{b}} &
\colhead{$E_{\rm SN}$\tablenotemark{c}} &
\colhead{$E_{\rm wind}$\tablenotemark{c}} & 
\colhead{Region} \\
\colhead{} &
\colhead{(M$_{\odot}$)} &
\colhead{(Myr)} &
\colhead{(M$_{\odot}$)} &
\colhead{} &
\multicolumn{4}{c}{($\times$10$^{53}$~ergs)} &
%\colhead{($\times$10$^{53}$~ergs)} &
%\colhead{($\times$10$^{53}$~ergs)} &
%\colhead{($\times$10$^{53}$~ergs)} &
%\colhead{}
}
\startdata
 1~~~~ & 5$\times$10$^{5}$ & 14.0~~~ & 13.9 &  4300~~ & 43~ & 4.8~ & 42.9~ & 15.1~ & ~~D03\\
 2~~~~ & 1$\times$10$^{5}$ &  7.0~~~ & 18.3 &  580~~ & 5.8~ & 0.96~ & 3.4~ & 3.0~ & ~~D01\\
 3~~~~ & 4$\times$10$^{5}$ &  6.0~~~ & 19.4 &  2100~~ & 21~ & 3.8~ & 10.0~ & 11.9~ & ~~D05\\
 4~~~~ & 2$\times$10$^{5}$ &  6.0~~~ & 19.4 &  110~~ & 11~ & 1.9~ & 5.0~ & 6.0~ & ~~D05\\
 5~~~~ & 1.5$\times$10$^{5}$ &  4.0~~~ & 22.9 &  610~~ & 6.1~ & 1.4~ & 1.0~ & 2.6~ & ~\nodata\\
 6~~~~ & 1.2$\times$10$^{5}$ &  14.0~~~ & 13.9 &  1000~~ & 10~ & 1.1~ & 10.3~ & 3.6~ & ~~D02\\
 7~~~~ & 1.2$\times$10$^{5}$ & 5.5~~~ & 20.1 &  600~~ & 6.0~ & 1.1~ & 2.5~ & 3.4~ & ~~D07\\
 8~~~~ & 1.5$\times$10$^{5}$ &  5.5~~~ & 20.1 &  750~~ & 7.5~ & 1.4~ & 3.1~ & 4.2~ & ~\nodata\\
 9~~~~ & 3$\times$10$^{5}$ &  3.5~~~ & 24.1 &  1100~~ & 11~ & 2.9~ & 0~ & 4.1~ & ~~D10\\
10~~~~ & 1.5$\times$10$^{5}$ &  13.0~~~ & 14.3 &  1200~~ & 12~ & 1.4~ & 11.9~ & 4.5~ & ~\nodata\\
11~~~~ & 5$\times$10$^{5}$ ? &  6 ?~~~ & 19.4 &   2600~~ & 26 ~ & 4.8~ &12.6~ &14.9~ & ~~D10\\
12~~~~ & 1$\times$10$^{5}$ &  6.0~~~ & 19.4 &  530~~ & 5.3~ & 0.96~ & 2.5~ & 3.0~ & ~~D02\\
13~~~~ & 7$\times$10$^{4}$ &  5.0~~~ & 20.9 &  330~~ & 3.3~ & 0.67~ & 1.1~ & 1.7~ & ~~D07\\
14~~~~ & 3$\times$10$^{5}$ &  3.0~~~ & 25.7 &  1000~~ & 10~ & 2.9~ & 0~ & 3.0~ & ~~D10\\
15~~~~ & 5$\times$10$^{4}$ &  7.5~~~ & 17.8 &  300~~ & 3.0~ & 0.48~ & 1.9~ & 1.5~ & ~\nodata\\
16~~~~ & 2$\times$10$^{5}$ & 14.0~~~ & 13.9 &  1700~~ & 17~ & 1.9~ & 17.1~ & 6.0~ & ~\nodata\\
17~~~~ & 1$\times$10$^{5}$ &  5.5~~~ & 20.1 &  500~~ & 5.0~ & 0.96~ & 2.1~ & 2.8~ & ~~D09\\
18~~~~ & 1.2$\times$10$^{5}$ &  4.0~~~ & 22.9 &  490~~ & 4.9~ & 1.1~ & 0.8~ & 2.1~ & ~~D07\\
19~~~~ & 6$\times$10$^{4}$ &  5.5~~~ & 20.1 &  300~~ & 3.0~ & 0.57~ & 1.2~ & 1.7~ & ~~D09\\
20~~~~ & 3$\times$10$^{5}$ &  3.0~~~ & 25.7 &  1000~~ & 10~ & 2.9~ & 0~ & 3.0~ & ~~D10\\
21~~~~ & 5$\times$10$^{4}$ &  5.5~~~ & 20.1 &  250~~ & 2.5~ & 0.48~ & 1.0~ & 1.4~ & ~\nodata\\
22~~~~ & 1.5$\times$10$^{4}$ &  7.5~~~ & 17.8 &  90~~ & 0.90~ & 0.14~ & 0.6~ & 0.5~ & ~~D01\\
23~~~~ & 8$\times$10$^{4}$ &  5.5~~~ & 20.1 &  400~~ & 4.0~ & 0.77~ & 1.6~ & 2.2~ & ~\nodata\\
24~~~~ & 7$\times$10$^{4}$ & 14.0~~~ & 13.9 &  610~~ & 6.1~ & 0.67~ & 6.0~ & 2.1~ & ~\nodata\\
25~~~~ & 1$\times$10$^{5}$ &  3.5~~~ & 24.1 &  380~~ & 3.8~ & 0.99~ & 0~ & 1.4~ & ~\nodata\\
26~~~~ & 3.5$\times$10$^{4}$ &  7.5~~~ & 17.8 &  210~~ & 2.1~ & 0.34~ & 1.4~ & 1.1~ & ~~D10\\
27~~~~ & 7$\times$10$^{4}$ & 14.0~~~ & 13.8 &  610~~ & 6.1~ & 0.67~ & 6.0~ & 2.1~ & ~\nodata\\
28~~~~ & 8$\times$10$^{4}$ &  5.5~~~ & 20.1 &  400~~ & 4.0~ & 0.77~ & 1.6~ & 2.2~ & ~\nodata\\
29~~~~ & 2$\times$10$^{4}$ &  8.0~~~ & 17.3 &  120~~ & 1.2~ & 0.19~ & 0.9~ & 0.6~ & ~~D01\\
30~~~~ & 7$\times$10$^{4}$ &  5.5~~~ & 20.1 &  350~~ & 3.5~ & 0.67~ & 1.4~ & 2.0~ & ~~D10\\
31~~~~ & 1$\times$10$^{4}$ &  7.5~~~ & 17.8 &  60~~ & 0.60~ & 0.096~ & 0.4~ & 0.3~ & ~~D01\\
32~~~~ & 5$\times$10$^{4}$ &  4.0~~~ & 22.9 &  200~~ & 2.0~ & 0.48~ & 0.3~ & 0.9~ & ~\nodata\\
33~~~~ & 7$\times$10$^{4}$ &  6.5~~~ & 18.8 &  390~~ & 3.9~ & 0.67~ & 2.1~ & 2.1~ & ~~D10\\
34~~~~ & 3$\times$10$^{5}$ &  6.0~~~ & 18.5 &  1700~~ & 17~ & 2.9~ & 7.5~ & 9.0~ & ~~D04\\
35~~~~ & 2$\times$10$^{4}$ &  7.5~~~ & 17.8 &  120~~ & 1.2~ & 0.19~ & 0.8~ & 0.6~ & ~~D08\\
36~~~~ & 3$\times$10$^{5}$ &  14.0~~~ & 13.9 &  2600~~ & 26~ & 2.9~ & 25.7~ & 9.1~ & ~\nodata\\
37~~~~ & 1.5$\times$10$^{4}$ &  7.5~~~ & 17.8 &  90~~ & 0.90~ & 0.14~ & 0.6~ & 0.5~ & ~\nodata\\
38~~~~ & 3$\times$10$^{5}$ &  6.5~~~ & 18.8 &  1700~~ & 17 ~ & 2.9~ & 8.9~ & 9.0~ & ~~D04\\
39~~~~ & 7$\times$10$^{4}$ &  5.5~~~ & 20.1 &  350~~ & 3.5~ & 0.67~ & 1.4~ & 2.0~ & ~~D10\\
40~~~~ & 5$\times$10$^{4}$ &  6.0~~~ & 19.4 &  260~~ & 2.6~ & 0.48~ & 1.3~ & 1.5~ & ~~D01\\
41~~~~ & 1.2$\times$10$^{4}$ &  7.5~~~ & 17.8 &  72~~ & 0.72~ & 0.11~ & 0.5~ & 0.4~ & ~~D07\\
42~~~~ & 1.5$\times$10$^{4}$ &  7.5~~~ & 17.8 &  90~~ & 0.90~ & 0.14~ & 0.6~ & 0.5~ & ~~D06\\
43~~~~ & 5$\times$10$^{4}$ &  14.0~~~ & 13.9 &  430~~ & 4.3~ & 0.48~ &  4.3~ & 1.5~ & ~~D03\\
44~~~~ & 5$\times$10$^{4}$ &  7.5~~~ & 17.8 &  300~~ & 3.0~ & 0.48~ & 1.9~ & 1.5~ & ~~D03\\
45~~~~ & 7$\times$10$^{4}$ &  6.0~~~ & 19.4 &  370~~ & 3.7~ & 0.67~ & 1.8~ & 2.1~ & ~\nodata\\
46~~~~ & 2$\times$10$^{4}$ &  6.5~~~ & 18.8 & 110~~ & 1.1 ~ & 0.19~ &0.6~ &0.6~ & ~~D01\\
47~~~~ & 4$\times$10$^{4}$ &  6.0~~~ & 19.4 &  210~~ & 2.1~ & 0.38~ & 1.0~ & 1.2~ & ~\nodata\\
48~~~~ & 8$\times$10$^{4}$ & 15.0~~~ & 13.5 &  720~~ & 7.2~ & 0.77~ & 7.3~ & 2.4~ & ~~D08\\
49~~~~ & 8$\times$10$^{4}$ &  6.0~~~ & 19.4 &  420~~ & 4.2~ & 0.77~ & 2.0~ & 2.4~ & ~\nodata\\
50~~~~ & 1$\times$10$^{4}$ &  7.5~~~ & 17.8 &  60~~ & 0.60~ & 0.096~ & 0.4~ & 0.3~ & ~~D02\\
\enddata
\tablenotetext{a}{Cluster numbers come from Table 1 of \citet{Whitmore99}.}
\tablenotetext{b}{Estimated using the method outlined in \S{5}.}
\tablenotetext{c}{From Starburst99 model.}
\end{deluxetable}

\newpage

\begin{deluxetable}{lrrc}
\tablecolumns{8}
\tablewidth{0pc}
\tablecaption{Upper Limits for Stellar Wind Energy}
\tablehead{
\colhead{Type} & 
\colhead{Mass} & 
\colhead{$v_\infty$} & 
\colhead{$E_{\rm wind}$} \\
\colhead{} &
\colhead{(M$_{\odot}$)} &
\colhead{(km s$^{-1}$)} &
\colhead{(ergs)}
}
\startdata
~O3 & 120~~ & 3,150~~ & 5.9$\times$10$^{51}$ \\
~O5 &  60~~ & 1,885~~ & 1.1$\times$10$^{51}$ \\
~O8 &  23~~ & 1,530~~ & 2.7$\times$10$^{50}$ \\
~B0 &  17.5 & 1,535~~ & 2.1$\times$10$^{50}$ \\
~B3 &   7.6 &   590~~ & 1.3$\times$10$^{49}$ \\
\enddata

\end{deluxetable}

\newpage

\begin{deluxetable}{crrcrrrc}
\tablecolumns{8}
\tablewidth{0pc}
\tablecaption{One-Component MEKAL Models with Solar and Free-Varying 
Abundances}
\tablehead{
\colhead{Region} & 
\multicolumn{3}{c}{Solar Abundance} &
\multicolumn{4}{c}{Free-Varying Abundances} \\
\colhead{} & 
\colhead{$kT$} &
\colhead{$\chi^2$/DoF} &
\colhead{$E_{\rm th}$} &
\colhead{$kT$} &
\colhead{$Z$} &
\colhead{$\chi^2$/DoF} &
\colhead{$E_{\rm th}$} \\
\colhead{} &
\colhead{(keV)} &
\colhead{} &
\colhead{(ergs)} &
\colhead{(keV)} &
\colhead{($Z_\odot$)} &
\colhead{} &
\colhead{(ergs)}
}
\startdata
 D01   & 0.37 & 126.7/28 = 4.5 & 1.5$\times10^{54}$ & 0.43 & 0.06 &  58.6/27 = 2.2 & 5.2$\times10^{54}$ \\
 D02   & 0.37 &  20.5/~8 = 2.6 & 4.3$\times10^{53}$ & 0.40 & 0.04 &   7.0/~7 = 1.0 & 1.6$\times10^{54}$ \\
 D03   & 0.58 &  36.2/13 = 2.8 & 1.7$\times10^{54}$ & 0.53 & 0.07 &   9.1/12 = 0.8 & 4.6$\times10^{54}$ \\
 D04   & 0.76 &  47.6/10 = 4.8 & 7.2$\times10^{53}$ & 0.78 & 0.13 &  23.8/~9 = 2.6 & 1.6$\times10^{54}$ \\
 D05   & 0.61 &  46.2/10 = 4.6 & 6.2$\times10^{53}$ & 0.64 & 0.05 &  11.3/~9 = 1.3 & 2.0$\times10^{54}$ \\
 D06   & 0.54 & 189.4/42 = 4.5 & 5.6$\times10^{54}$ & 0.53 & 0.10 &  88.7/41 = 2.2 & 1.4$\times10^{55}$ \\
 D07   & 0.62 & 121.5/21 = 5.9 & 2.0$\times10^{54}$ & 0.61 & 0.08 &  59.0/20 = 2.9 & 5.5$\times10^{54}$ \\
 D08   & 0.58 & 144.4/33 = 4.4 & 3.6$\times10^{54}$ & 0.58 & 0.10 &  61.4/32 = 1.9 & 9.4$\times10^{54}$ \\
 D09   & 0.62 &  47.9/~6 = 8.0 & 4.8$\times10^{53}$ & 0.71 & 0.02 &   8.8/~5 = 1.8 & 1.9$\times10^{54}$ \\
 D10   & 0.67 &  82.9/22 = 3.8 & 1.5$\times10^{54}$ & 0.68 & 0.12 &  28.6/21 = 1.4 & 3.6$\times10^{54}$ \\
 D11   & 0.66 &  93.5/38 = 2.5 & 2.2$\times10^{54}$ & 0.67 & 0.21 &  51.7/37 = 1.4 & 4.4$\times10^{54}$ \\
\enddata
\end{deluxetable}

\newpage

\begin{deluxetable}{crrcrrrc}
\tablecolumns{8}
\tablewidth{0pc}
\tablecaption{One- versus Two-Component MEKAL Models with Solar Abundance}
\tablehead{
\colhead{Region} & 
\multicolumn{3}{c}{One-Component} &
\multicolumn{4}{c}{Two-Component} \\
\colhead{} & 
\colhead{$kT$} &
\colhead{$\chi^2$/DoF} &
\colhead{$E_{\rm th}$} &
\colhead{$kT_1$} &
\colhead{$kT_2$} &
\colhead{$\chi^2$/DoF} &
\colhead{$E_{\rm th}$} \\
\colhead{} &
\colhead{(keV)} &
\colhead{} &
\colhead{(ergs)} &
\colhead{(keV)} &
\colhead{(keV)} &
\colhead{} &
\colhead{(ergs)}
}
\startdata
 D01   & 0.37 & 126.7/28 = 4.5 & 1.5$\times10^{54}$ & 0.22 & 0.63 &  52.7/26 = 2.0 & 2.8$\times10^{54}$ \\
 D02   & 0.37 &  20.5/~8 = 2.6 & 4.3$\times10^{53}$ & 0.18 & 0.56 &   7.0/~6 = 1.2 & 7.1$\times10^{53}$ \\
 D03   & 0.58 &  36.2/13 = 2.8 & 1.7$\times10^{54}$ & 0.19 & 0.64 &   9.8/11 = 0.9 & 2.2$\times10^{54}$ \\
 D04   & 0.76 &  47.6/10 = 4.8 & 7.2$\times10^{53}$ & 0.70 & 6.6  &   8.3/~8 = 1.0 & 8.1$\times10^{54}$ \\
 D05   & 0.61 &  46.2/10 = 4.6 & 6.2$\times10^{53}$ & 0.57 & 3.4  &   8.6/~8 = 1.1 & 4.9$\times10^{54}$ \\
 D06   & 0.54 & 189.4/42 = 4.5 & 5.6$\times10^{54}$ & 0.23 & 0.61 & 119.1/40 = 3.0 & 7.5$\times10^{54}$ \\
 D07   & 0.62 & 121.5/21 = 5.9 & 2.0$\times10^{54}$ & 0.23 & 0.70 &  79.8/19 = 4.2 & 2.7$\times10^{54}$ \\
 D08   & 0.58 & 144.4/33 = 4.4 & 3.6$\times10^{54}$ & 0.32 & 0.85 &  80.6/32 = 2.5 & 5.7$\times10^{54}$ \\
 D09   & 0.62 &  47.9/~6 = 8.0 & 4.8$\times10^{53}$ & 0.56 & 6.6  &   9.8/~4 = 2.5 & 8.8$\times10^{54}$ \\
 D10   & 0.67 &  82.9/22 = 3.8 & 1.5$\times10^{54}$ & 0.61 & 1.5  &  40.7/20 = 2.0 & 3.8$\times10^{54}$ \\
 D11   & 0.66 &  93.5/38 = 2.5 & 2.2$\times10^{54}$ & 0.63 & 3.5  &  48.5/36 = 1.3 & 1.1$\times10^{55}$ \\
\enddata
\end{deluxetable}

\newpage

\begin{deluxetable}{crlrcrlrc}
\tablecolumns{9}
\tablewidth{0pc}
\tablecaption{Two-Component MEKAL versus APEC with Solar Abundance}
\tablehead{
\colhead{Region}                         & 
\multicolumn{4}{c}{Two-Component MEKAL} &
\multicolumn{4}{c}{Two-Component APEC} \\
\colhead{}             & 
\colhead{$kT_1$}       &
\colhead{$kT_2$}       &
\colhead{$\chi^2$/DoF} &
\colhead{$E_{\rm th}$} &
\colhead{$kT_1$}       &
\colhead{$kT_2$}       &
\colhead{$\chi^2$/DoF} &
\colhead{$E_{\rm th}$} \\
\colhead{}       &
\colhead{(keV)}  &
\colhead{(keV)}  &
\colhead{}       &
\colhead{(ergs)} &
\colhead{(keV)}  &
\colhead{(keV)}  &
\colhead{}       &
\colhead{(ergs)}
}
\startdata
 D01   & 0.22 & 0.63 &  52.7/26 = 2.0 & 2.8$\times10^{54}$ & 0.24 & 0.70 &  51.3/26 = 2.0 & 2.8$\times10^{54}$ \\
 D02   & 0.18 & 0.56 &   7.0/~6 = 1.2 & 7.1$\times10^{53}$ & 0.18 & 0.57 &   3.7/~6 = 0.7 & 7.5$\times10^{53}$ \\
 D03   & 0.19 & 0.64 &   9.8/11 = 0.9 & 2.2$\times10^{54}$ & 0.18 & 0.64 &   6.1/11 = 0.6 & 2.3$\times10^{54}$ \\
 D04   & 0.70 & 6.6  &   8.3/~8 = 1.0 & 8.1$\times10^{54}$ & 0.41 & 0.97 &  34.6/~8 = 4.3 & 1.1$\times10^{54}$ \\
 D05   & 0.57 & 3.4  &   8.6/~8 = 1.1 & 4.9$\times10^{54}$ & 0.59 & 2.6  &   7.5/~8 = 0.9 & 3.8$\times10^{54}$ \\
 D06   & 0.23 & 0.61 & 119.1/40 = 3.0 & 7.5$\times10^{54}$ & 0.21 & 0.61 &  85.7/40 = 2.1 & 7.7$\times10^{54}$ \\
 D07   & 0.23 & 0.70 &  79.8/19 = 4.2 & 2.7$\times10^{54}$ & 0.22 & 0.69 &  62.6/19 = 3.3 & 2.7$\times10^{54}$ \\
 D08   & 0.32 & 0.85 &  80.6/32 = 2.5 & 5.7$\times10^{54}$ & 0.33 & 0.80 &  66.3/32 = 2.1 & 5.8$\times10^{54}$ \\
 D09   & 0.56 & 6.6  &   9.8/~4 = 2.5 & 8.8$\times10^{54}$ & 0.59 & 4.9  &   8.9/~4 = 2.2 & 6.7$\times10^{54}$ \\
 D10   & 0.61 & 1.5  &  40.7/20 = 2.0 & 3.8$\times10^{54}$ & 0.25 & 0.76 &  52.9/20 = 2.6 & 2.0$\times10^{54}$ \\
 D11   & 0.63 & 3.5  &  48.5/36 = 1.3 & 1.1$\times10^{55}$ & 0.65 & 2.1  &  36.6/36 = 1.0 & 6.9$\times10^{54}$ \\
\enddata
\end{deluxetable}

\end{document}